\documentclass[12pt,aps,prd,reprint,superscriptaddress, nofootinbib,showpacs,preprintnumbers,floatfix,longbibliography]{revtex4-1}
\usepackage{amsmath, latexsym, amssymb, hyperref, graphicx, color, multirow}
\usepackage[table]{xcolor}
\usepackage{tabularx}
\usepackage[T1]{fontenc}
\usepackage[capitalize]{cleveref}
\usepackage{booktabs} 
\usepackage{fouriernc} 
\definecolor{nicered}{rgb}{.7,.1,.1}
\definecolor{nicegreen}{rgb}{.1,.5,.1}
\definecolor{darkblue}{rgb}{0,0,.5}
\hypersetup{colorlinks, citecolor=nicegreen,linkcolor=nicered, urlcolor=darkblue}
\newcommand{\beq}{\begin{equation}}
\newcommand{\eeq}{\end{equation}}
\newcommand{\beqa}{\begin{eqnarray}}
\newcommand{\eeqa}{\end{eqnarray}}

\begin{document}

\title{Neutrino: chronicles of an aloof protagonist \footnote{Based on a series of colloquia by G.S. during the year 2019 in the US, China and Taiwan.}}

\author{Alejandra Melfo} 
 
\affiliation{Centro de F\'isica Fundamental, Universidad de Los Andes,
M\'erida, Venezuela}

\author{Goran Senjanovi\'c}

\affiliation{%
Arnold Sommerfeld Center, Ludwig-Maximilians University,  Munich, Germany 
}%

\affiliation{%
International Center for Theoretical Physics,
Trieste, Italy}

\date{\today}

\begin{abstract}
We give a brief  account of the history of neutrino, and how that most aloof of all particles has shaped our search for a theory of fundamental interactions ever since it was theoretically proposed. We introduce the necessary concepts and phenomena in a non-technical language aimed at a physicist with some basic knowledge of quantum mechanics.  In showing that neutrino mass could be the door to new physics beyond the Standard Model, we emphasize the need to frame the issue in the context of a  complete theory, with testable predictions accessible to present and near future experiments. We argue in favor of the Minimal Left-Right Symmetric theory  as the strongest candidate for such theory, connecting neutrino mass with parity breakdown in nature. This is the theory that led originally to neutrino mass and the seesaw mechanism behind its smallness, but even more important,  the theory that sheds light on a fundamental question that touches us all: the symmetry between left and right. 

\end{abstract}

\maketitle

\section{Prelude}

The story of the neutrino is a tale of scientific breakthroughs and human drama, of missed opportunities and the occasional good luck. The story we tell here is also a personal  account of the struggle  to come up with a self-contained, predictive theory of neutrino mass, analogous of the Standard Model of electromagnetic and weak interactions. Self-contained here means not needing any additional assumption - the characteristic of true theories of nature. This will become clear in the process of recapitulating the salient features of the Standard Model.
%as a theory of the origin of the charged fermion and gauge boson masses. 

Our aim here is to convince the reader that such a theory not only has been found, but it has been there for a long time: the well-known Left Right Symmetric theory. And that it is the same theory that led to the prediction of a non-zero neutrino mass and the celebrated seesaw mechanism behind its smallness, long before experiment. Originally suggested to account for parity (transformation from left to right) violation in the weak interactions, it turned to be a theory of the origin of neutrino mass - just like the Standard Model, that started as a theory of weak interactions and turned into a theory of the origin of  masses of all elementary particles (but the neutrino). In recent years it emerged that the Left Right theory is completely self-contained and predictive, again just like the Standard Model itself.

The Standard Model, otherwise an extremely successful high precision theory, predicts massless neutrino by its structure, paving the way for going beyond.  Neutrino mass is thus today our best candidate for a door into new physics, and arguably the Large Hadron Colliders, or its successor, could open that door. 

Before we start, one clarifying comment. This review is written in the spirit of physics department colloquia that G.S. gave during the year 2019 in the US, China and Taiwan.  He was invited  by the editors of Modern Physics Letters A to publish it, and was joined by A.M. who had participated actively in the preparation of these colloquia. The starting colloquium given at Fermilab during the Neutrino summer school in August of 2019, was actually a hybrid between a general public talk and a talk for physicists, and thus had to be even more pedagogical. We believe thus that the material here, which follows the spirit of that talk, ought to be comprehensible to anyone  with  some working knowledge of quantum mechanics.

We hope though that even a high energy specialist may find it of some use, providing them with a sense of perspective and outlook, if nothing else. If she wishes, though, to get a sense of perspective without going through this somewhat long and probably too pedagogical essay, we suggest a recent short outlook of the question of the origin of neutrino mass, based on the Neutrino 2020 Closing Theory Talk~\cite{Senjanovic:2020rcq}.

\section{Elementary particles}

It is often said that the world is essentially formed by quarks  and the electron. After all, two kinds of quark, up and down, conspire to make the protons and neutrons in atomic nuclei,  which together with the electron conspire in turn to make atoms - which make you and us and all matter we know.  However, this picture is actually not completely true: the interaction energy of quarks can also be considered a constituent of matter as described below, and interactions in turn have associated particles. Schematically, on the other hand, quarks suffice to explain all the properties of nuclei in terms of their quantum numbers such as say electromagnetic charge.

The proton and neutron are the best known members of a larger family, called the hadrons, from the greek word {\it hadros}: stout, thick. It takes two up and one down quark to form the proton, and two down and one up for the neutron.  
Just like protons and neutrons, quarks are fermions, particles with spin $s = 1/2$. So is the electron, the most familiar member of the class of particles called leptons, from the greek {\it lepton}: thin, fine. All fitting, as the electron mass is about one thousandth of the neutron or proton masses, which are approximately equal. 

Like electrons, quarks are very light, with up and down quarks having a mass equivalent to a few times the electron's mass, more than a hundred times smaller than the masses of proton and neutron. Since the mass of the atoms is concentrated mostly in their nuclei, and the quark masses are basically negligible,  this means most of the mass in matter comes from the kinetic energy of quarks, forever confined inside the hadrons. 
Almost all of your mass comes not from mass itself, but from the energy of motion: a surprising fact but yet another confirmation that mass and energy are basically one and the same concept, as the Special Theory of Relativity would have it.

Although the world that we see and touch is made up of quarks and electrons, the main protagonist of this saga is the elusive neutrino, another, very different lepton, one that has captivated the imagination of physicists ever since its existence began to be suspected. What makes it so special is its aloofness. It carries no electromagnetic charge, like the neutron; but, unlike the neutron, it does  not even  have an electromagnetic dipole moment as far as we know. But moreover, neutrinos snub everything they pass through on their endless journey to the edge of the Universe. A neutrino produced in a reactor has a mean free path (the distance it can travel before interacting with another particle) of around $10^{20} cm$, some ten million times the distance from the Earth to the Sun, thousands of light years. For an electron produced in the same reactor, it is smaller than a centimetre. The indifference of neutrinos is truly amazing.

It is not surprising that such a  particle has been a window into new physics. Physicists love it for its coolness. In 1960, John Updike
begged to disagree in his poem {\it Cosmic Gall}

\begin{verse}
 {\it NEUTRINOS, they are very small.  \\
 They have no charge and have no mass  \\
 And do not interact at all. \\
The earth is just a silly ball \\
To them, through which they simply pass,  \\
 Like dustmaids down a drafty hall  \\
 Or photons through a sheet of glass.  \\
 They snub the most exquisite gas, \\
  Ignore the most substantial wall,\\
    Cold shoulder steel and sounding brass,  \\
    Insult the stallion in his stall, \\
And scorning barriers of class, \\
     Infiltrate you and me! Like tall \\
and painless guillotines, they fall  \\
     Down through our heads into the grass.  \\
     At night, they enter at Nepal \\
and pierce the lover and his lass \\
      From underneath the bed -- you call \\
It wonderful; I call it crass.}
 \end{verse}
 
 It is remarkable that the poet finds the neutrino aloofness crass, and that it provides a cosmic irritation to him, while we do find it wonderful -- after all, this is what makes neutrino such a good window into new physics.   At the time Updike wrote his poem, the neutrino was indeed thought to be massless, so he took it for granted -- this tells you how important the issue was and still is. In fact, neutrino is not massless, but it is incredibly light, a least a million times lighter than its sibling the electron.

\begin{figure}[h]
\centerline{
\includegraphics[width=.9\columnwidth]{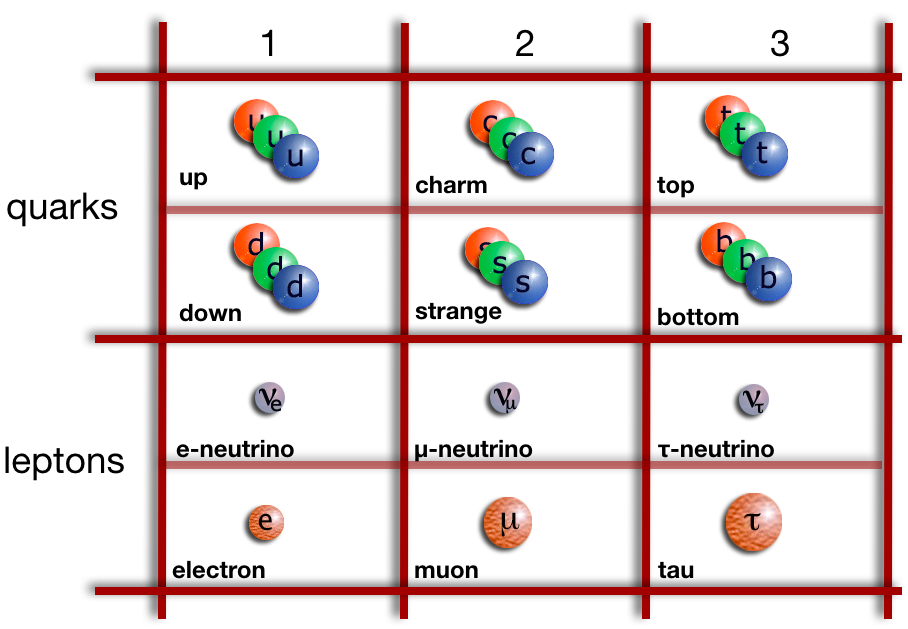}  }
\caption{Quarks and leptons come in three generations, exact replicas of the lightest ones. The only difference are the values of the masses. }
\label{fig-generations}
\end{figure}

We have then a nicely ordered  symmetric world consisting  of four elementary particles:  two quarks (up and down) and two leptons (electron and neutrino). Quarks combine in threes to form baryons, hadrons with $s = 1/2$, the notable representatives being proton and neutron. They also combine in twos to form another kind of composite particles, the spin zero mesons. Leptons go their way alone, the electron completing the atoms, the neutrinos traveling indifferently around. But this, of course,  is not all. These four  elementary  particles have replicas, particles with exactly the same characteristics, but (at least in the case of quarks and electrons, and probably also for neutrinos), with larger masses. We call these replicas generations, and there are three of them, pictured in Fig. \ref{fig-generations}.

Particles in the other generations - except for neutrinos -  are short lived, decaying quickly into their first generation versions, which justifies our claim that the universe is made out of first generation fermions only.  But this means that there are interactions between the generations, it is said that the generations ``mix''. They do it in an orderly fashion: quarks mix with quarks and leptons with leptons, and the mixings respect the next-neighbour pattern. The extra generations would not be important at all in this tale, were it not for the existence of three types of neutrinos, called electron, muon and tau neutrinos ($\nu_e,\nu_\mu, \nu_\tau$). They are all too light  to decay faster than the age of the universe, which makes them stable for all practical purposes. Thus they are still present today, filling the universe quietly.

This fact that there is more than one generation  is indispensable for the existence of the phenomenon of {\it neutrino oscillations},  for now the only experimental manifestation of neutrino mass, as we will see below.  To see how this comes about, we need to introduce the crowning success of 20th century Quantum Field Theory: the gauge theory of particle physics, a prelude to the Standard Model of particle interactions. 

\section{The Standard Model of particle interactions}\label{SMsection}

\subsection{Quantum numbers: charges, colors and flavors}\label{charges}

\begin{figure}[h]
\centerline{
\includegraphics[width=.2\columnwidth]{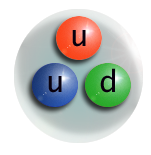}\hskip 1cm   \includegraphics[width=.2\columnwidth]{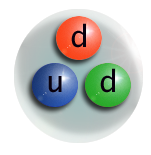}  }
\centerline{proton \hskip 2cm neutron}
\caption{Protons and neutrons are composed of different combinations of three up and down quarks.}
\label{fig-pn}
\end{figure}

We have talked about the constituents of matter and their fancy names, but what really distinguishes one elementary particle form the other? How can we tell that the invisible object in our detector is, say, an electron? Simply: it behaves like one. In other words, an elementary particle is defined by its quantum numbers (mass, charge, spin...), which in turn dictate how they interact. Take for example the up and down quarks, that combine in threes to form the proton and neutron, as in Figure \ref{fig-pn}.

   By convention, in our units, the electric charge of the electron is $q_e = -1$. The neutrality of atoms implies then the proton charge to be unit ($q_p=+1$), and the neutron to be neutral ($q_n=0$). This in turn implies that quarks have fractional charges
$$
q_u = 2/3 , \qquad q_d = -1/3 .
$$
Notice that these charges seem fractional only due to our desire to keep the historically chosen unit of charge to be $q_e = - 1$, the electron charge.
The unit of electromagnetic charge is  $e$, and it defines the electromagnetic coupling between elementary particles, to be discussed below.  As common in the field of particle physics, hereafter we use the so-called natural units $\hbar = c = 1$.

Incidentally, the familiar charges of neutrons, protons and electrons are either zero or integer multiples of the unit of the smallest charge $q_d$, to an incredible precision of 1 in $10^{20}$. This is the phenomenon of charge quantization, and amazingly enough it holds true for all known particles, be they elementary or composite. 
This miracle would be explained by the existence  in our Universe of magnetic monopoles (we have only seen magnetic charge  in the form of dipoles, or magnets, up to now). Namely, in a quantum mechanical theory with even just one monopole in the Universe, the product of electric and magnetic charge must be an integer for the sake of consistency, as showed by Dirac~\cite{Dirac:1948um}. Interestingly, in a theory built on charge quantization magnetic monopoles must exist~\cite{tHooft:1974kcl, Polyakov:1974ek}, which completes the deep connection between charge quantization and monopoles.  That is another story, however, and as beautiful as it is, it is outside the scope of our tale here. 

Quarks have an additional, kind of picturesque quantum number, a new charge that comes in three different kinds. This additional quantum number, called the color (because just like with real colors, you need three of them to make a neutral, or white, object), makes it possible for three up  quarks to be confined inside a hadron, for example -- a configuration that would be forbidden by Pauli exclusion principle otherwise.  In Fig. \ref{fig-generations}, the color charge is represented by painting the three quarks green, blue and red. A crucial role is played by  yet another charge, the so-called flavor (or weak isospin). Up and down quarks are just different flavors according to this nomenclature, as are neutrino and electron, the same is true for particles in the other generations.

The existence of quarks was suggested independently in 1964 by George Zweig and Murray Gell-Mann \cite{Zweig:1981pd, Zweig:1964jf, GellMann:1964nj}, as an attempt to classify the emerging zoo of countless baryons and mesons. Gell-Mann, by that time a professor in Caltech and an undisputed leader in the particle physics world, stopped short of taking them seriously as physical entities and attached basically only mathematical importance to them, a way of book-keeping. Zweig, also at Caltech but at the time only a graduate student, on the contrary, showed a great belief in their existence and went on to write a masterpiece as a postdoctoral research associate at CERN. However, nobody believed in quarks - after all, Gell-Mann, the guru himself, did not - and so Zweig's paper ended up unpublished (as can be seen in the references). To this day he has not received a proper recognition for his great work, the core of the field of elementary particles.

Other characteristics of the particles are inferred by their interactions (or the absence of them). For example, quarks have a baryon number $B$, leptons a lepton number $L$, as we will see below. Both baryon and lepton number appear conserved in nature, just like the electromagnetic charge -- but there is a strong suspicion that it is only a matter of time before the violations of these apparent laws are observed. 

The important thing though is that particles are identified by their masses and spin, and a set of numbers depending on the way they interact with each other: the electromagnetic charge, the color charge and the flavor charge. These charges are conserved at the fundamental level, and as we will see in section VII, they correspond to global symmetries of the theory. Thus the Standard Model of elementary particles is a model of their interactions. Not all of them: only the ones strong enough to be relevant at the quantum level. This excludes paradoxically  the gravitational interaction, the best known of forces and the one that governs the large scale behavior of our Universe. The gravitational force between two protons is some 38 orders of magnitude weaker than the electromagnetic force between them, and it only counts when you have a very large number of protons and neutrons (the Sun has some $10^{60}$). 

\subsection{Fundamental forces and messengers }\label{forces}

Therefore the Standard Model concerns itself with three interactions, also called forces:
\begin{itemize}

\item {\it Electromagnetic}, described in the framework of Quantum Electro Dynamics (QED), a quantum field theory based on the $U(1)$ gauge group.

\item {\it Strong}, described in the framework of Quantum Chromo Dynamics (QCD), a quantum field theory based on the $SU(3)$ gauge group.

\item {\it Weak}, described in the framework of Quantum Flavor Dynamics, a quantum field theory based on the  $SU(2)$ gauge group.

\end{itemize}

\begin{figure}[h]
\centerline{
\includegraphics[width=.7\columnwidth]{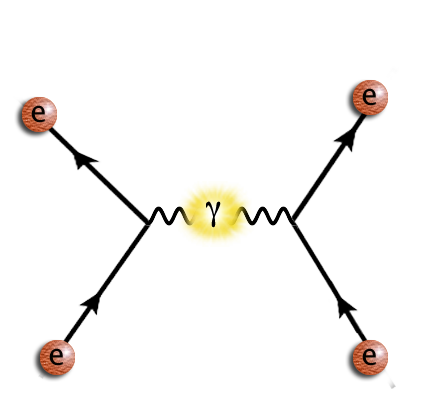}  }
\caption{Electron-electron scattering is mediated by the photon, ($\gamma$), the messenger of the electromagnetic interaction.}
\label{fig-qed}
\end{figure}

In Quantum Field Theory, these forces or interactions are mediated by messengers. Messengers are not fermions as matter particles, but bosons with spin=1, and are called the gauge bosons. The most familiar example is the photon, the particle of light. In QED, the electromagnetic force between two particles such as two electrons, is the result of one of these electrons emitting a virtual photon, and the other absorbing it. In 1948, Richard Feynman \cite{Feynman:1948km} devised a practical way of representing this via the so-called Feynman diagrams, as in Figure \ref{fig-qed}. These diagrams are much more than a pictorial representation, they are in fact a powerful tool enabling physicists to perform complicated and difficult calculations of scattering amplitudes.   They are simultaneously extremely helpful for seeing what goes on physically - ideal for our purpose of illustration. Fermions are represented by straight lines with arrows, and the messengers by wavy lines connecting them. 
 
 The interaction term in the electromagnetic Hamiltonian is
\beq \label{Hqed}
{\cal H}_{QED} = e \bar f Q_{em} \gamma^\mu f A_\mu\,.
\eeq
That is, fermions $f$ interact with each other according to their charge $Q_{em}$ in units of the electron's charge $e$, through  the interchange of a photon $A_\mu$. In these units, $Q_{em}$ is simply a collection of the above discussed electric charges, normalised to the  electron charge $q_e = - 1$. The function $f$ describing a fermion is actually a four-component spinor, a relativistic generation of the concept of spinors in Quantum Mechanics.
%, while the electromagnetic coupling is defined as $\alpha = e^2/ 4 \pi$.
The reader who is not familiar with this formula and wishes to have a short summary of the relativistic theory of spinors, and the Dirac $\gamma_\mu$ matrices, should consult the Appendix.

\begin{figure}[h]
\centerline{
\includegraphics[width=.9\columnwidth]{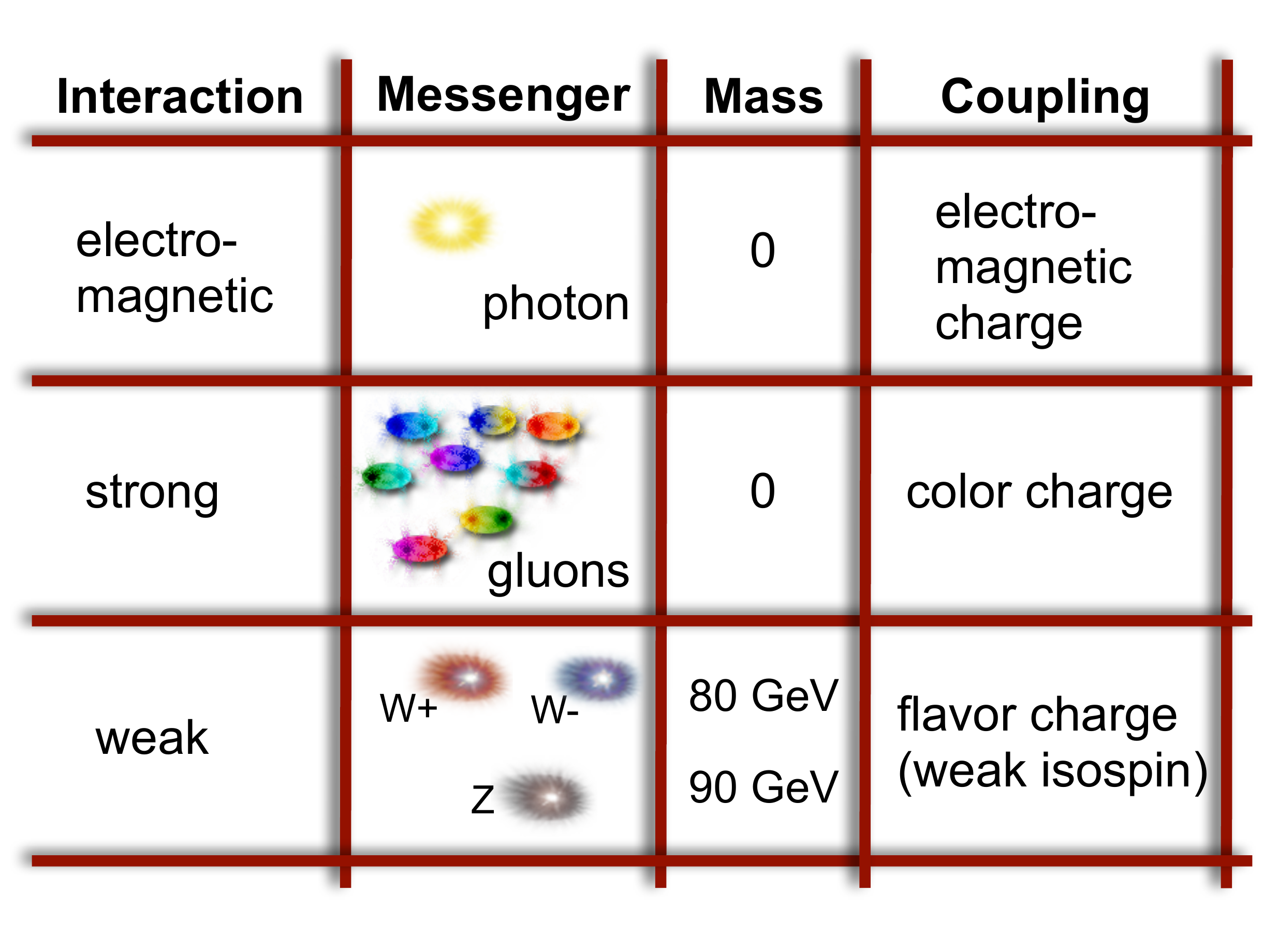}  }
\caption{The three  interactions relevant in particle physics at present energies (that is, excluding gravity), and their messengers.  }
\label{fig-interactions}
\end{figure}

The interpretation of the electromagnetic force as an interaction mediated by the photon in the early 20th century was extremely successful and led to the great achievements of QED. With time it  became clear that other interactions could be understood this way, paving the way for the Standard Model. These fundamental interactions are summarized in Figure \ref{fig-interactions}. 

The messengers of the strong force, an interaction occurring only between quarks, are called gluons and there are eight of them, corresponding to eight generators of the $SU(3)$ group. They are also bosons and also massless, just like the photon - but they carry color charge as the quarks (thus the name Quantum Chromodynamics).  Unlike the photon, by having color they interact among themselves and that makes all the difference. While the electromagnetic coupling becomes smaller as the distance between charged particles increases, allowing them to go freely their own ways, the strong coupling keeps growing with the distance. And there is convincing evidence that it grows so much that it never allows quarks to be free - an isolated quark has never been seen in Nature - confining them permanently into baryons and mesons. The phenomenon of confining is fascinating, and while it still  lacks a rigorous proof in the context of the theory, it has a firm computational and observational basis.

When the distance decreases,  or equivalently  (according to the uncertainty principle) the energy increases, the strong coupling becomes weaker and weaker~\cite{Politzer:1973fx,Gross:1973id}. This implies that at energies much bigger than the proton mass - or distances much smaller than the proton Compton radius - the quarks should appear basically free. Thanks to this phenomenon, known as {\it asymptotic freedom},  quarks  have been 'observed'  in the deep inelastic scattering of electrons on protons. At such high energies the hadrons, both baryons and mesons, behave just as if made of lightly bound, almost free, quarks.  The composite particles formed by quarks are like your  shopping bag; as you walk, or need be even run, the contents  may move almost freely inside, but they will not escape if your bag is tightly closed.
 The crucial point is in the word {\it almost}; quarks do interact among themselves  in accord with the QCD theory. Today, at least at high energies, QCD has been confirmed experimentally to a great precision. 
 
When it comes to the weak force, however, things change radically. The messengers, the W and Z bosons, are massive to start with, making their interactions short-ranged. This seemingly innocent fact leads to the weak interaction being very special, connected to the fascinating phenomenon of Spontaneous Symmetry Breaking and the Higgs mechanism, as we shall see later.   As the energy increases, beyond the W and Z boson masses, the weak interaction ceases to be weak since the basic coupling is similar in strength to the electromagnetic one. The only reason they appear so weak at low energies, much below W and Z masses, is the inertia of the messengers, the W and Z bosons. At today's energies strong interactions are not strong and weak interactions are not weak.

The weak force is also the only interaction felt by the neutrino, and it is deeply connected with its discovery.  We need to step back to the beginnings, long before the Standard Model was developed.

\section{Weak interactions and the birth of neutrinos}

\subsection{Beta decay: predicting neutrino}\label{nubeta}

There is yet another feature that makes neutrino very special: it was  predicted by theory beforehand. It all started with the well-known radioactive process dubbed beta decay, whereupon an atom of  (say) cobalt (27 protons, 33 neutrons) is observed to decay into an atom of nickel (28 protons, 32 neutrons), emitting the so-called beta radiation. Beta radiation is in fact a beam of electrons, and the process was later understood to consist of a neutron  in the cobalt nuclei decaying into a proton and an electron (the neutron is just sufficiently heavier than the proton so that this can take place)
\beq
n \longrightarrow p + e\,.
\eeq

 However, be it that we talk of nuclei or neutrons and protons, conservation of energy dictates that such a process would result on the electron having a definite energy, and not a continuous spectrum, as was observed.  In 1930, Wolfgang Pauli came up with what he called ``a desperate remedy" to save the law of conservation of energy. Unable to personally communicate his idea to his colleagues in T\"ubingen because his presence in a ball in Z\"urich was ``indispensable", he wrote a famous letter \cite{Pauli:1930pc}  to his ``Dear Radioactive Ladies and Gentlemen". He suggested that a fourth particle had to be involved in beta decay, so that it was the sum of its energy plus that of the electron that had a definite value. Such particle had to be neutral , and be a fermion with spin=1/2 just as the electron. 
He called it logically a neutron, since the true neutron was not yet discovered.

 It was only after Chadwick discovered neutron in 1932 that the neutrino got its Italian name, ``little neutral one".  Story has it that  it was Amaldi,  one of the students in Enrico Fermi's group in Rome, who came up with this historic nomenclature in an effort to differentiate it from the recently discovered neutron. 

With the experimental data available to Pauli, it was already known that the new particle had to have a small mass, perhaps of order the mass of the electron. Later, more precise beta decay data further lowered this upper limit, and neutrinos are now known to have a mass less than one electronvolt in our natural units, or one millionth of an electron mass.  Nowadays, when many of us in particle physics enthusiastically propose the existence of a plethora of new particles in our models, it is difficult to imagine the deep dilemma Pauli confronted when proposing the neutrino. Not only physicists were far more cautious then, there was also the question of detecting such a weakly interacting, tiny particle. Pauli even came to regret introducing a ghost particle that would seemingly never be seen. But we have said neutrinos almost never interact, and the point again is in the word "almost". It is true that you will not discover a singular neutrino, but if you have many of them, your chances of winning the neutrino lottery increase statistically. 

  The solution turned out to be simple: use nuclear reactors that produce huge numbers of neutrinos, on the order of ten trillions per cm squared per second. Using the Savannah River reactor, Reines and Cowan  devised an experiment whereby neutrino hits a proton in the water and creates a neutron and positron. A relatively small detector of some 100 liters sufficed to produce tens of events per hour. It took some time, but finally in 1956 they could confirm the existence of neutrinos \cite{Cowan:1992xc}. Upon receiving their telegram announcing the great discovery, 
Pauli  famously said: "everything comes to him who knows how to wait". 

\subsection{The weak interaction}\label{fermi}

Beta decay is the first example discovered of a process caused by the weak interaction. Fermi, the father of the theory  of weak interactions, described this process in what is today called the effective Fermi theory.  The interaction among neutrons $n$, protons $p$, electrons $e$ and neutrino $\nu$ is described  at low energies (of the order of proton mass), by~\cite{Fermi:1934hr} by the Hamiltonian
 \beq
 {\cal H}_{eff} = \frac{G_F}{\sqrt{2}}  (\bar p \gamma_\mu n)   (\bar e \gamma^\mu \nu)\, ,
 \eeq
where $G_F \simeq 10^{-5} m_p^{-2}$ is the effective coupling.  It is the original example of an effective field theory, the approach that helps you understand the phenomenology of a new force even before having an underlying, fundamental model of the messenger in question.

We will jump over years of discoveries in physics and a rich history, and introduce the modern view of what goes on. For a neutron to become a proton, it is enough that one of its down quarks decays into an up quark, plus an electron and an anti-neutrino
\beq
d \longrightarrow u + e + \bar \nu\, .
\eeq
  This process not only conserves energy and momentum, but also electric charge (as the reader can check), and baryon and lepton numbers. Baryon number ($B$) is defined so that quarks have  $B=1/3$ (there are three quarks in a baryon) and leptons $B=0$. Conversely, leptons have lepton number $L=1$, and quarks $L=0$. 
 We will introduce anti-particles later, for now suffice to say that an anti-lepton (here the antineutrino) has $L=-1$. That the weak interactions (and in fact, the others as well) preserve baryon and lepton numbers, means that quarks cannot be converted into leptons or {\it vice versa}. 
 
\begin{figure}[h]
\centerline{
\includegraphics[width=.7\columnwidth]{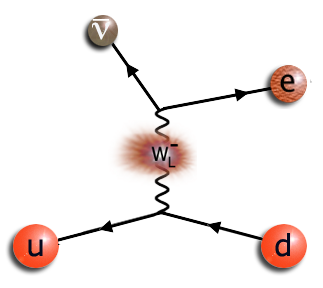}  }
\caption{In beta decay, a $d$ quark becomes an $u$ quark by emitting a virtual $W$ boson, which in turn decays in an electron and an anti-neutrino. }
\label{fig-betadecay}
\end{figure}

  In the modern view, weak interactions as the beta decay of $d$ are mediated by gauge bosons, $W$ in this case, as pictured  in Fig. \ref{fig-betadecay}.

%\subsection{Fundamental theory of weak interaction}\label{fund}

The interaction Hamiltonian for this process looks now similar to the QED one we saw in Eq.\ref{Hqed}
\beq\label{Hbeta}
{\cal H}_{weak} = \frac{g}{2 \sqrt{2}} \left[\bar u \gamma^\mu (1+\gamma_5) d + \bar \nu \gamma^\mu (1+ \gamma_5)  e\right] \, W_\mu^+ + h.c.\, ,
\eeq
where $g \simeq e$ measures the strength of the weak interaction.

The factor $(1 + \gamma_5)$ accounts for a fundamental feature of weak interactions, parity violation, that we will discuss later; for the time being the reader can just ignore it.  The picture is actually more complicated because there are more generations, and we remarked before that the generations mix. We can be now more precise: they mix in their interactions with the $W$ boson, and only in these interactions.  For example, the interaction of $d$ with $u$ above is in fact an interaction of $d$ with $u$, $c$ and $t$, with different strengths. This is best represented in {\it generation space}, where each flavor of quarks and leptons is a three-dimensional vector, and their interactions are given by unitary 3x3 unitary mixing matrices. 

For example, in the quark sector 
\beq\label{mixings}
{\cal H} = \frac{g}{2 \sqrt{2}}  \left (\bar u, \bar c, \bar t \right) \gamma^\mu (1+ \gamma_5) V_{\rm CKM} \left(\begin{array}{c} d \\ s \\ b \end{array} \right) \, W_\mu^+ + h.c.\,.
\eeq
The  mixing matrix  $V_{\rm CKM}$ is called the Cabibbo-Kobayashi-Maskawa matrix~\cite{Cabibbo:1963yz},
while its leptonic analog $V_{\rm PMNS}$ is called the Pontecorvo-
Maki-Nakagawa-Sakata matrix~\cite{Pontecorvo:1957qd}, in  honour of their fathers. 
The former matrix is associated with the prediction of Kobayashi and Maskawa that there be the third generation of quarks and leptons for which they got Nobel prize. Sadly,  in one of the greatest scientific injustices ever, the prize did not include Zweig, the true father of the quark picture. 
%Time flies but there is still hope that the Nobel committee corrects its great sin.
It is the PMNS matrix, however, that is of fundamental importance for us, for it leads to the fascinating physics of neutrino oscillations. You do not need to know much about these 
mixings, except that the three quark mixing angles are much smaller than their leptonic counterparts.

The fundamental (gauge) theory of weak interactions, analogous to QED theory of electromagnetism, was to be developed much later by Glashow, Weinberg and Salam, in the 1960s. It predicted not only the existence of three messengers (the $W^+, W^-$ and $Z^0$ gauge bosons) and the form of their interactions, but also the exact ratio of their masses. They were discovered in 1983 in the  Super Proton Synchrotron at CERN,  the first large modern circular collider with a 7 km circumference. The weak force messengers are far from massless, the $W$  is 80 times as heavy as the proton, and the $Z$ is 90 times as much, in perfect agreement with the theory. 

In what follows we will be using the physical picture of messengers, even when speaking of distant history before their existence was suggested. The reason is simple - this makes everything much more clear. 

Although Pauli had suggested that the neutrino was very light, later experimental upper limits were even smaller than he envisioned. And the minimal Standard Model, the Glashow-Weinberg-Salam theory of weak and electromagnetic interactions~\cite{Glashow:1961tr,Weinberg:1967tq,Salam:1968rm}, was eventually built with a strictly massless neutrino. However, Italian physicist Bruno Pontecorvo always believed  that neutrino does have have a mass, just like its sibling electron.  He noted that if neutrinos have a mass, however tiny, and with even only two generations of them, their quantum mechanic nature could make them change one species into another as they travel through space \cite{Pontecorvo:1969ys} - as long as the their mass difference was  non-vanishing but sufficiently small.

\subsection{Neutrino oscillations}\label{oscillations}

To illustrate this, imagine a simpler world with only two generations of neutrinos. If they are massive, neutrinos mix. What we call an electron neutrino (namely, the neutrino that interacts with the electron) will be produced in the Sun as a coherent state, a linear combination of two particles, $\nu_1$ and $\nu_2$, with different masses
\beq
\nu_e = \cos\theta \nu_1 + \sin\theta\nu_2\,,
\eeq
and what we call a muon neutrino will be the orthogonal combination
\beq
\nu_\mu = -\sin\theta \nu_1 + \cos\theta\nu_2\,.
\eeq
While traveling, each state $\nu_1, \nu_2$ propagates differently, as they have different mass. Therefore the  coefficients of the linear combinations change in time, and it is possible to calculate the probability of $\nu_e$ becoming a $\nu_\mu$ because of this
\beq
P(\nu_e \to \nu_\mu) = \sin^22\theta \sin^2 \frac{\Delta m^2 L}{4 E}\,,
\eeq
where $L$ is the distance traveled, $E$ the energy and $\Delta m^2 $ is the difference in mass squared, between $\nu_1$ and 
$\nu_2$. 

 The effect is known as an ``oscillation", with neutrinos changing flavors back and forth.  However, it  is not that one neutrino particle oscillates into another. Rather, a coherent state of different physical particles is produced originally, and if physical particles have different masses, the original state will evolve differently. If the mass difference is small, it will remain a coherent superposition of the physical states. In this sense we can see where the above formula comes from. 
 
 The probability of the oscillation is maximal when the mixing is maximal, which is clearly $\theta = \pi/4$ ($\theta = 0$ means no mixing and $\theta = \pi/2$ is simply a redefinition of this situation) - hence the $\sin^2 2\theta$ term. The oscillation probability must be proportional to the mass difference (neutrinos with equal mass are basically the same species) and with the distance. But in the limit of infinity energy, the mass difference becomes irrelevant which explains the  $1/E$ dependance. Except for the factor of 4 in the numerator,  the oscillation formula can be ``derived'' on physical and dimensional grounds only.
 
And thus, if neutrinos oscillate, an electron neutrino produced by ordinary, first-generation matter in the Sun, will metamorphose as it travels to Earth. It may then arrive at our detector in the form of a muon neutrino, and since our detector is designed to detect electron neutrinos, the muon neutrino would be invisible to it. 
Pontecorvo argued that a neutrino detector could then see a deficit of neutrinos coming form the Sun, that is, less than expected according to the stellar model the astrophysicists were probing for our star. The same would happen with the so-called atmospheric neutrinos, those produced high in the Earth's atmosphere by the incidence of cosmic rays.   His work was, however, largely ignored. 

   The point is that Pontecorvo was a communist, deluded with Western society, and in the 1950 he became the protagonist of an unusual and rather dramatic tale. He escaped to the Soviet Union in a spectacular fashion,  with his wife and three kids hidden in the trunk of a car. It was the middle of the cold war era, and he was not forgiven by the West for a long time. It appears that he was never fully trusted in the East either, and finding himself in a kind of no man's land did not do well for his career as a physicist. Amazingly enough, when in the seventies and eighties it was becoming more and more clear that Pontecorvo could be right - there were already indications of the solar neutrino deficit as predicted by oscillations - the West still ignored his work and talked of a solar neutrino ``puzzle''. 
      
 Pontecorvo died prematurely, before he could witness the proof of his ideas. His dramatic life resembles the tale of solar neutrino oscillations, which reads like a detective story. Indeed the story of his life, in the background of ever present neutrino, is beautifully depicted in a book by Frank Close appropriately entitled {\em Neutrino} ~\cite{Close:2010zz}.

 That neutrinos do indeed oscillate, at least the atmospheric ones,  was confirmed finally in 1998 by the Kamioka Nucleon Decay Experiment. It is the muon neutrino produced in pion decays that oscillates into the tau neutrino in this case, with a probability that implies a mass difference of order of the tenths of $eV$, some $10^{-7} $ times the electron's mass.
 And the solar neutrino oscillations were confirmed three years later by the Sudbury Observatory experiment, this time implying an oscillation between electron and muon neutrinos, with a mass difference roughly an order of magnitude smaller.

 Unfortunately, neutrino oscillations cannot tell us what the neutrino mass is, only the mass difference. 
  Moreover, neutrino mass can be very different from charged particles masses. To see how this comes about, we need to turn to yet another piece of fascinating history, involving the discovery of antimatter, and an even more mysterious tale about another great  Italian physicist - this time a disappearance that was never resolved.

\section{Relativistic fermions and Dirac's equation}

\subsection{Helicity and chirality}

We have said that particles are labeled by their quantum numbers, which in turn indicate the way they react to the forces (electromagnetic charge, color charge, flavor charge), or other characteristics such as its mass and spin. The spin of a particle, its intrinsic angular momentum, determines also its statistics, namely how it behaves collectively: either as a fermion if the spin is a semi-integer (subject to Pauli's exclusion principle, which forbids two fermions to occupy the same quantum state), or gregariously as a boson if it is an integer. Up to now, we have only detected elementary fermions with spin 1/2, and bosons with spin 1 (the messengers) or 0 (the Higgs particle).  

In order to represent a particle in  Quantum Field Theory,  mathematical properties are assigned to the fields for each of these labels. Thus a charged particle will be represented by a complex field, a ``colored'' quark as a three-component vector of the group $SU(3)$ (it is not essential for the reader the precise meaning of the $SU(3)$ group) and so on. And all quarks and leptons, as we already remarked, have to be  {\em spinors}, as Quantum Mechanics dictates for particles with spin 1/2. However they cannot be the usual 2-component spinors, for they have to obey also Special Relativity, as we now discuss.

If particles are massless, one can define its {\it helicity}, the projection of its spin in the direction of its motion.
Notice that this a relativistically invariant quantity. A particle moving in the direction $z$, with its spin  in the same direction $z$,  has positive helicity; one moving in the same direction with spin in the opposite ($-z$) direction has negative helicity. So we have by convention two kinds of spinors: 
a left-handed $u_L$ with helicity $h = -1/2$, and right-handed $u_R$ with the opposite helicity, $h = 1/2$.

Helicity looses its meaning for a massive particle, since motion is not a relativistically invariant notion.  We can always stop the particle, making helicity vanish in its rest frame  - this is why for massive fermions we only speak of spin, defined precisely when the particle is at rest. 
We can for example go to the rest frame and look at the state with positive spin the $z$ direction, $s_z = 1/2$.  If we now  look at it from a reference frame moving in the $+ z$ direction, we see it has a negative helicity, $ h = -1/2$. This would have to be represented by a left-handed two-component spinor $u_L$. However, we can also go to the coordinate frame moving in the opposite direction, and then its helicity would be positive, corresponding to a right-handed spinor $u_R$. 

We see that while a massless particle has a fixed helicity, and is represented with a two-component (Weyl) spinor,  the massive  one needs  two such states for a proper relativistic description. The two states $u_L$ and $u_R$ are distinct, and instead of helicity they have {\it chirality}, a property that cannot be eliminated by going to the rest frame.

  Chirality  is an intrinsic property, just like mass or spin, that can be thought of as the particle ``handedness". An electron is in fact made of two states: the right handed  electron $e_R$, and the left handed  electron $e_L$. The left and right states are two component spinors, and together they form the {\it four}-component spinor. Left and right-handed particles are mirror images of each other, and this mirror symmetry we call parity (or left-right symmetry). 

We leave the technical details to the Appendix, but we can now better describe  equation \ref{Hbeta}. The terms  $1 + \gamma_5$ are projectors, picking out one of the chirality states out of a four-component spinor. The weak interaction Hamiltonian in Eq. \ref{Hbeta} involves only the left-handed electron $e_L = 1/2 (1 + \gamma_5) e $, and if we perform a parity transformation on it, it would not look the same. This fundamental property of the weak interaction, that it does not look the same in the mirror, will be central to our tale.

 \subsection{Antiparticles}

 In 1928, Paul Dirac set out to write down the analog of Schr\"odinger equation, but valid for a fast moving, relativistic electron~\cite{Dirac:1928hu}. Arguably modern particle physics started then, with the equation  that bears his name
\beq \label{dirac}
( i \gamma^\mu \partial_\mu - m ) \, \psi = 0\,.
\eeq 
Here $\psi$ describes a relativistic particle with mass, and as we have said and Dirac masterfully demonstrated, it has to be a complex, four-component spinor field as discussed in the Appendix. 

Dirac immediately realized an unexpected consequence:  His equation simultaneously described an electron and a new particle, with the same mass but opposite charge (not only electromagnetic, but whatever other charge it could possibly have) - an anti electron. This was inevitable, once $\psi$ contained two distinct chiral states. The four-component spinor represents the particle, but has inside the information of an antiparticle. A mathematical transformation, charge conjugation, performed on the spinor gives us a different combination of the chiral states: an antiparticle. 

A positively charged electron, or positron, had not been observed at the time and was in fact so unthinkable that Dirac was reluctant to ascribe it a physical reality, preferring to think about it as a construct of theory. It is hard not to have a longing for that era, when claims mattered and a physicist's reputation depended on them, now that it seems like anything goes and a new model gets proposed almost daily. It took Dirac three years to finally say it clear and loudly. 
The positron was discovered indeed, soon after, in  cosmic rays~\cite{Anderson:1932zz}, when a highly energetic photon pair produces an electron and a positron. 
It was a bombshell. 

Dirac's equation is moreover  valid for any massive fermion. Not only the world of fermions was suddenly duplicated, a particle in contact with its own anti-particle would annihilate into pure energy in the form of photons. The fascinating concept escaped the physics labs, and soon books and movies were full of antimatter weapons annihilating everything in sight. In 1955, Segr\'e and Chamberlain completed the picture by discovering the anti-proton\cite{Chamberlain:1955ns}, and today every charged particle has its own anti-particle. 

\begin{figure}[h]
\includegraphics[width=.9\columnwidth]{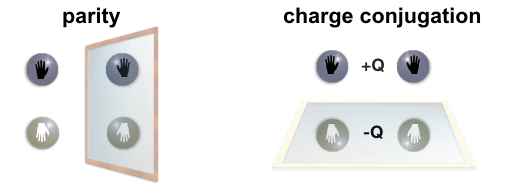}
\caption{Two of Nature's mirror symmetries and their actions over chiral states. Parity interchanges chiralities, charge conjugation inverts all charges.}
\label{fig-mirrors}
\end{figure}

The connection between chirality and antiparticles is profound.  Anti-particles have exactly opposite chiralities from their corresponding particles - for example, an anti-particle of $e_L$ is right-handed. Just as the parity transformation  turns left into right and vice versa, the charge conjugation - the transformation that changes a particle into its anti-particle and vice versa - is a mirror symmetry, as represented in Fig.\ref{fig-mirrors}.
Thus, instead of using $e_L$ and $e_R$, one can speak of  a left-handed electron and a left-handed positron (or the right-handed counterparts) - or simply of an electron and a positron.

That was then the picture provided by Dirac's equation: matter is made up of fermions, and for each fermion there is an anti-fermion with the same mass but with all charges of opposite sign. However, less than ten years later a  young Italian physicist discovered an exception to the rule: the electrically neutral fermions, such as neutrino. His name was Ettore Majorana, and his story is among the most mysterious and fascinating in the history of science. 

\subsection{Majorana particles}

In the words of Enrico Fermi:  ``There are many categories of scientists, people of second and third rank, who do their best, but do not go very far. There are also people of first class, who make great discoveries, which are of capital importance for the development of science. But then there are the geniuses, like Galilei and Newton. Well, Ettore was one of these. Majorana had greater gifts that anyone else in the world; unfortunately he lacked one quality which other men generally have: plain common sense" \cite{fermiquote}. For example, it was almost impossible to get him to publish even Nobel prize caliber work. 

In March 1938, soon after he started lecturing at Naples University, 32-years-old Majorana disappeared during a boat trip from Palermo to Napoli and his body was never found. Several hints suggested the explanation was suicide, but this could not be confirmed and the legend started: Majorana was officially considered disappeared and speculations about his fate (secluded in a monastery in Tuscany?) abounded. Some years ago,  the {\it Procura di Roma} revised evidence and admitted the possibility that he had escaped for unknown reasons to Valencia, in Venezuela, living under a different name until his death.  Whatever happened, fortunately Majorana left us a  legacy, and it has to do with the special way he obtained his professorship in Naples.

 Since his publishing record was so scarce, the condition for the position was that he produced a paper - and Majorana took out of a drawer a work that would mark the field of neutrinos ever after. In that paper \cite{Majorana:1937vz}, the last before his disappearance, he showed that a fermion can be in principle its own anti-particle, or more precisely, a state made up half and half of particle and anti-particle, later called a Majorana fermion. This is possible only for a completely neutral particle, and only if it has mass. If such is the nature of the neutrino, it can be tested in processes that violate electron (in general lepton) number, producing two electrons ``out of nothing" instead of the usual pair electron-positron. One of this processes is a nuclear decay without neutrinos in the final state (neutrinoless double beta decay), but there is also the interesting possibility of an analog high-energy collider process, that could be even under the reach of the Large Hadron Collider (LHC). But let us first see how this may come about.

\subsection{ Two kinds of mass}\label{twomass}

\begin{figure}[h]
\includegraphics[width=.7\columnwidth]{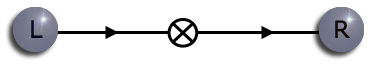}
\caption{Dirac mass which connects otherwise independent left-handed and right-handed spinors.}
\label{fig-diracmass}
\end{figure}

The reason we do not often hear about two different versions of a charged particle is that the left and right versions of chiral fermions, $f_L$ and $f_R$, are connected by their mass term
\beq
{\rm Dirac \, mass}:  m_D \, f_R^\dagger f_L\,,
\eeq
which is represented in a Feynman diagram by a mass insertion, shown in Fig \ref{fig-diracmass}. The convention here is that dagger (basically complex conjugation) on $f_R$ implies that it is going out, while $f_L$ is entering.  
A left-handed particle becomes a right-handed particle as it propagates, or in other words, they are the same object. 
The four-component Dirac spinor that we call the electron is the sum of the two spinors with fixed handedness $e = e_L + e_R$.
Only if the fermion is massless we can speak of a particle with fixed chirality (that is in that case its helicity), and then a single state suffices, either left-handed or right-handed.

\begin{figure}[h]
 \includegraphics[width=.7\columnwidth]{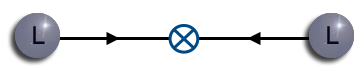}
\caption{Majorana mass connects particles of a given chirality with their antiparticles.}
\label{fig-majoramamass1}
\end{figure}

Majorana observed that another mass term is possible, involving only one of the two chiralities and its antiparticle,  today known as the Majorana mass term 
\beq
{\rm Majorana \, mass}:  m_M \, f_L^T i \sigma_2 f_L\,,
\eeq
where $\sigma_2$ is a purely imaginary antisymmetric Pauli matrix. This expression is invariant under Lorentz symmetry, since by anti-symmetrising two $s=1/2$ states one gets a spin zero state. By the above convention, on both ends $f_L$ is entering, indicating the violation of whatever charge this state may carry. It is represented in a Feynman diagram accordingly, as in Fig. \ref{fig-majoramamass1}

 In fact, Dirac's equation also says that an antiparticle moving forwards in time can be interpreted as a particle moving backwards in time. In a Feynman diagram, this means that the line for a fermion can have its arrow inverted, and represent instead a line for an antifermion. So the Majorana mass insertion diagram can also be seen with an antiparticle coming out, 
 and can be interpreted as a particle of a given chirality propagating and becoming an antiparticle (with the opposite chirality). In this sense the Majorana mass term is basically the same as the Dirac one, but with a particle changing into an anti-particle.  This is of course only possible if the particle has no electromagnetic charge, otherwise it would have to change the sign of the charge on flight. Any charge that such a particle were to carry would be broken by two units, which is why only a  neutral particle can have such a mass term.

Thus the particle with Majorana mass is its own antiparticle, and just as in the case of a massless particle, it can be described by a single two-component (Weyl) spinor. Such a fermion is then called a Majorana fermion.

\begin{figure}[h]
\includegraphics[width=.95\columnwidth]{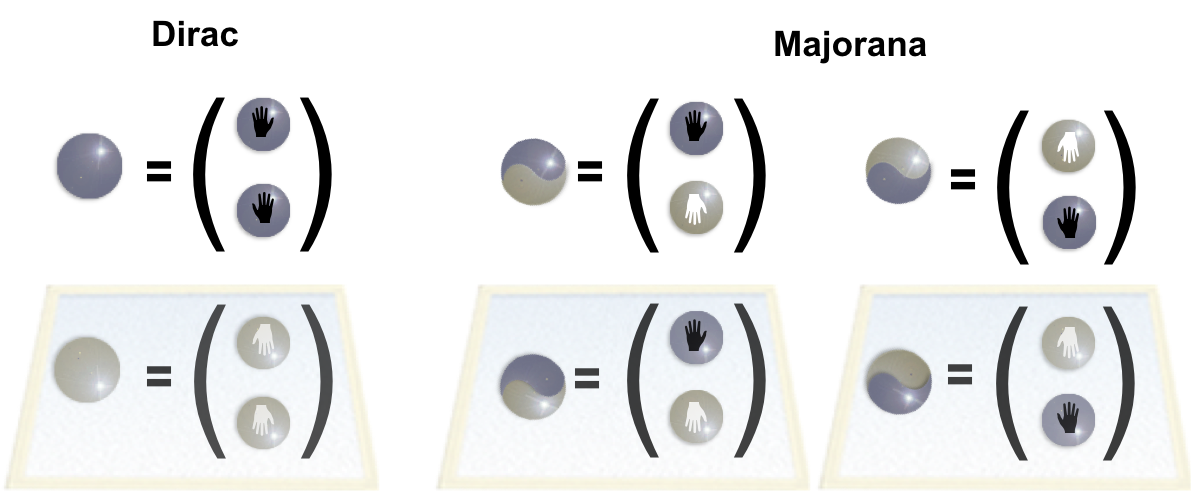}
\caption{A Dirac spinor is made up of two distinct chiral states, its image in the charge-conjugation mirror is an anti-particle. Majorana spinors are constructed with one chiral state and its charge-conjugated image, so they look the same in the mirror.}
\label{fig-spinors}
\end{figure}

  But we have argued that a relativistic massive particle requires a four component spinor, possessing both chiralities - so how does the Majorana spinor go around it? Easy: it uses the fact that anti-particles have opposite chiralities from their corresponding particles. The Majorana four-component spinor is a hybrid  made of a particle and its anti-particle, and this way it has both chiralities when you include its anti-state.
A Dirac spinor, instead, describes an electron which has a different anti-particle, of opposite charge, as represented in Fig.\ref{fig-spinors}

We see now how this is related to neutrino: it is the only chargeless fermion we have, and therefore the only candidate for having a Majorana mass, instead of (or in addition to)  a Dirac mass. 

Neutrinos carry  lepton number though, just like electrons, and thus if they were Majorana spinors, this would  result in the creation (or disappearance) of two units of lepton number - in other words, the violation of electron number. Lepton number violating process have not been observed in nature, and this could be the telltale signature of the existence of a Majorana particle. Let us see first how this could come about in a nuclear process, and leave the collider analog  for later.

In 1935, Maria Goeppert-Mayer pointed out  \cite{GoeppertMayer:1935qp} that some  nuclei can experience double beta-decay when the usual beta decay is energetically forbidden.  For example, suppose a neutron in the nucleus of a germanium atom beta-decays. The element with one more proton and one less neutron is arsenic, but it turns out to be heavier than the original nucleus, so the process is not allowed. However, selenium which has two protons more (and two neutrons less) is lighter than germanium and thus two neutrons of germanium beta-decay simultaneously, producing a selenium nucleus, plus two electrons and two antineutrinos
\beq
^{76} Ge \to ^{76} Se + e + e + \bar \nu_e + \bar \nu_e\,.
\eeq
This is of course much more unlikely to happen than normal beta decay. While ordinary beta decay, for example cobalt into nickel, has a lifetime $\tau_\beta \simeq 5 yr$, double beta has $\tau_{2\beta} \simeq 10^{21} yr$. Experimental confirmation of Goeppert-Mayer's idea had to wait until 1985.

\begin{figure}[h]
\includegraphics[width=.7\columnwidth]{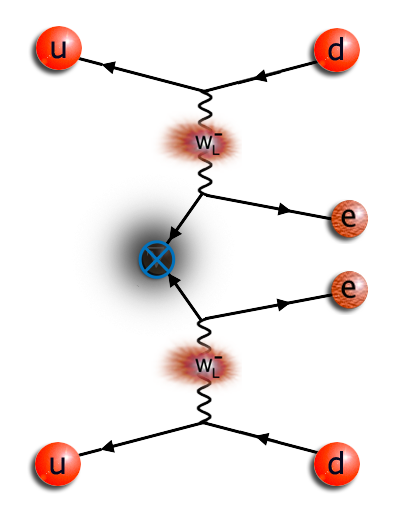}
\caption{Neutrinoless double-beta decay, with two down quarks giving rise to two up quarks and two electrons.}
\label{fig-doublebeta}
\end{figure}

However, if neutrino is its own antiparticle - a Majorana fermion - it is possible for both antineutrinos in double-beta decay to find each other and  kind of 'annihilate' through the Majorana mass term, as suggested in the late 30's by Racah and Furry \cite{Racah:1937qq,*Furry:1939qr} and illustrated in Fig. \ref{fig-doublebeta}. 
This is the so-called neutrinoless double beta decay ($0\nu2\beta$). The ugly name does not do it justice --this is a magical process, implying the violation of lepton number by two units. It is being actively searched for in tens of experiments all over the world. Its non-observation sets a strong lower bound, it turns out to be even more unlikely than double-beta decay with $\tau_{0\nu2\beta} \geq 10^{26} yr$~\cite{Agostini:2020xta}. This in turn implies that Majorana neutrino masses have an upper bound, $m_\nu^M \lesssim 0.1 eV$.

We see then that neutrino has yet another unique feature among the fermions: it can have in principle two {\em kinds} of mass. As we will see now, the nature and the origin of neutrino mass is related to the very foundations of the Standard Model. 

\section{Parity violation and the origin of mass}

\subsection{ A left-right asymmetric world} \label{pviolation}

In 1956, another bombshell was about to shake the world of physics. Cheng-Ning Yang and Tsung-Dao Lee  ~\cite{Lee:1956qn} asked an heretic question: what if left-right symmetry is broken in Nature? It was a radical proposal; up to then, every physics process was assumed to be left invariant under parity transformation - change left with right and vice-versa, and the new ``mirror image'' process will be exactly the same. Lee and Yang suggested than maybe the weak interaction was different in this sense. The idea was immediately put to test  by Chien-Shiung Wu \cite{Wu:1957my} who found not only that they were right, but that the left-right symmetry is actually maximally broken. This astonishing finding, confirmed by Leon Lederman's group in a simultaneous publication, \cite{Garwin:1957hc}, earned the two young physicists a Nobel prize already a year later - so dramatic was the impact of their proposal. Nature is not left-right symmetric at the particle level when the weak interaction is at play.

\begin{figure}[h]
\includegraphics[width=.7\columnwidth]{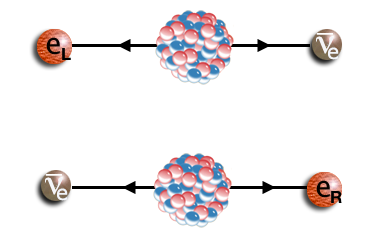}
\caption{Parity violation in $W^-$ decay. The outcome represented by the lower diagram is not observed.}
\label{fig-parityviolation}
\end{figure}

The situation is illustrated in Figure \ref{fig-parityviolation}. It involves the beta decay of a nucleus of cobalt, with spin 5, into nickel, with spin 4, with the emission of an electron (spin 1/2) and an  antineutrino (spin 1/2)
\beq
^{60}_{27} Co \to ^{60}_{26} Ni + e + \bar \nu_e\,.
\eeq

The nuclei are at rest and their spin points towards the right of the picture. Momentum conservation mandates that the electron and antineutrino escape in opposite directions, but angular momentum conservation implies that the spin of both electron and neutrino point in the same direction, to the right. Because the nuclei are  so heavy, electrons and neutrinos can both be considered to be massless particles to a very good approximation in this case, and we can identify their chirality with their helicity. Therefore, if the decay produces an electron moving towards the left, since its spin points in the opposite direction it has negative helicity, and it must be a left-handed electron $e_L$ (upper diagram in Fig. \ref{fig-parityviolation}). On the contrary, electrons escaping towards the right (lower diagram in Fig. \ref{fig-parityviolation}) are necessarily right-handed, $e_R$. Neutrinos are not detected in this experiment, but it is easy to check that they have the same chirality as the accompanying electron (being antiparticles).
 This way, the experimental setup connects the intrinsic, invisible property of chirality, with the left and right directions of the macroscopic world. Left-handed electrons go to the left, and right handed electrons go to the right (one could of course arrange the whole setup in the opposite way and make it more confusing, but physically equivalent).

What is observed is that all electrons produced in the decay go to the left.  The decay produces only $e_L$. The right-handed electrons do not participate in processes mediated by the weak interaction at all: parity is maximally violated in Nature. 

It is hard to overstate the impact of this finding. Parity, the symmetry between left and right, the first that a child learns looking at her image in the mirror, had always been assumed to hold true at the elementary level. For many, its violation was difficult to accept. Lee and Yang themselves, in the same paper where they postulated parity violation, wonder if there could perhaps exist ``corresponding elementary particles exhibiting the opposite asymmetry such that in the broader sense there will still be over-all right-left symmetry''. Once the experimental confirmation of their theoretical proposal came out, they seemed to have abandoned that idea.

\subsection{Fundamental theory of weak interactions} \label{fundweak}

Once parity breakdown was established, it took almost no time to deduce the form of the effective Fermi theory. A year later, Marshak and Sudarshan~\cite{Sudarshan:1958vf}, and Feynman and Gell-Mann~\cite{Feynman:1958ty},  came up with what is known as the $V-A$ theory.  Its central feature is the requirement that only left-handed quarks and leptons participate in weak interactions.

The Standard Model of particle interactions was then built on  the $V-A$ theory. Developed by Glashow, Weinberg and Salam over several years starting in 1961, it was based on the fundamental principle of gauge invariance just as QED before, and thus required the existence of new gauge bosons acting as messengers of the weak force. Another, equally crucial ingredient, was maximally broken parity: only left-handed particles can interact with the $W$ boson. In Weinberg's own words: 'V-A was the key'~\cite{Weinberg:2009zz}.

Thus the chiral nature of fermions becomes all-important. What we call an up quark is in fact two particles, one that has color and electromagnetic interactions only, another that in addition has weak interactions.  This is  taken into account in Eq. \ref{Hbeta} above when we wrote the weak interaction Hamiltonian, by the presence of the term $(1+\gamma_5)$, which has the effect of projecting the left-handed part of the spinors representing the different fermions. We can now write the weak Hamiltonian in terms of the left-handed fermions
\beq\label{Hbeta2}
{\cal H} = \frac{g}{4 \sqrt{2}} \left[\overline u_L \gamma^\mu d_L + \overline \nu_L \gamma^\mu  e_L \right] \, W_\mu^+ + h.c.\,.
\eeq

\begin{figure}[h]
\includegraphics[width=.9\columnwidth]{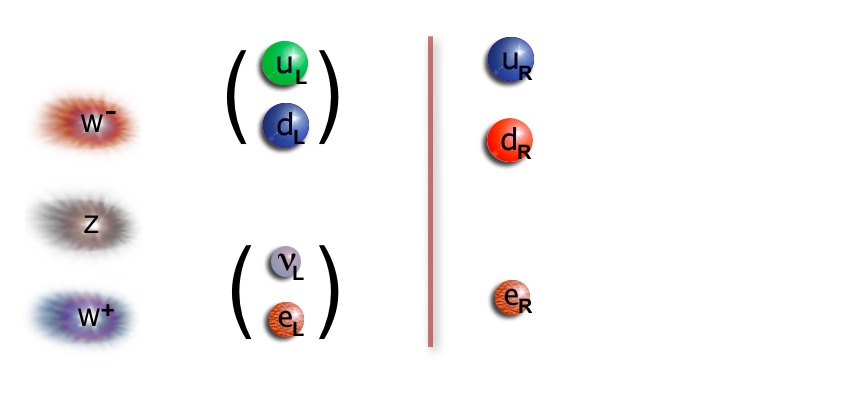}
\caption{The Standard Model has maximally broken parity, left-handed particles feel the weak force while right-handed ones do not. Here only quarks and leptons of the first generation are shown, together with the three messengers of the weak force. }
\label{fig-SM}
\end{figure}

In Figure \ref{fig-SM} the picture provided by the Standard Model for the first generation of fermions is shown. Parenthesis indicate that the up and down left-handed quarks  $u_L$ and $d_L$ are part of a doublet of the weak group $SU(2)$, which means that the weak interaction can change one into the other, as we have seen in the example of beta decay. The leptons $\nu_L$ and $e_L$ make up another such doublet, while their right-handed counterparts do not. The pattern is repeated in the other two generations.

Notably missing is the right-handed neutrino. The only interactions felt by neutrinos are weak interactions, and because parity is violated maximally, this interaction involves only the left-handed neutrino. Without an experimental evidence for a right-handed neutrino there is no need to have one in the Standard Model of particle interactions.

But of course, something has become deeply wrong with our picture. A weakly interacting left-handed particle cannot become a right-handed one while propagating, and suddenly stop feeling the weak force. The weak force, in violating parity, has forbidden all fermions to have a mass: the Dirac mass term is not invariant. As represented in Fig. \ref{fig-SM}, the Standard Model is a theory involving only massless particles. Enter the Higgs boson, the hero of 21st century experimental particle physics.

\section{The power of symmetry: The Standard Model}

\subsection{Symmetries, conserved charges and interactions}

Parity, the transformation of an object into its mirror image, is perhaps the best known of symmetries. But it is just one example  of the paramount importance of symmetries in physics, especially in particle physics. We have said that a particle is identified by its mass, spin, and all the set of charges that enable it to feel the different forces. These charges are conserved, just as energy and momentum are, in all processes. The connection between conserved charges and symmetry transformations was established by Emmy Noether and became one of the fundamental pillars of physics. Briefly, Noether's theorem states that for every global continuous symmetry of the theory there is a corresponding conserved quantity. 

This is a very deep concept. Suppose you perform a spatial rotation in your reference frame: components of vectors get mixed up, but the theory describes exactly the same phenomena. This invariance of the theory is connected to a conserved quantity, angular momentum. It works also for internal symmetries: interchange the color charges of all quarks and gluons, and your theory remains invariant, meaning that color charge is conserved. This is what we meant above when we said that QCD  is based on the symmetry group $SU(3)$, and QED on $U(1)$. The symmetry of the theory, what can be changed without it varying in its predictions, is enough to define how it works. Interactions are connected to charges, and charges to symmetries, so a theory of particle interactions becomes a theory of internal symmetries.

The symmetry transformations are expressed mathematically as phase transformations on the functions representing the particles, such as the one we encountered in Dirac's equation, the function $\psi$ representing a fermion.  We will not go into this rather technical detail, suffice it to say here that the crucial step is to have the phase transformations become local, i.e. space-time dependent. This has a profound immediate impact: if the theory is invariant under the phase transformation  but with  the phase depending on where we find ourselves, someone has to communicate this information to us - hence the need for a messenger of what becomes an interaction. This is the principle of local gauge invariance, the celebrated example being the gauge invariance of Maxwell's equations of electrodynamics. 

The power of symmetry in Quantum Field Theory was put to work in developing QED, and resulted in the picture of the interaction being the  interchange of a massless messenger, the photon. In a similar way, invariance under $SU(3)$ means that the strong force is mediated by eight massless bosons, the gluons. When developing the theory of weak interactions, Glashow, Weinberg and Salam built it around the symmetry group $SU(2)$, predicting three additional  messengers, the bosons $W^+,W^- $ and $Z$. These messengers  would seemingly have to be massless according to the same principle of gauge symmetry that gives massless gluons and the photon. 
The charge associated to the $SU(2)$ symmetry is  weak isospin. Up and down quarks form a weak isospin doublet, they are like components of a vector, and saying that the theory is invariant means that one can interchange up and down left-handed quarks. Or left-handed neutrinos and electrons, which also form a doublet: it is the equivalent in internal space of changing the reference frame. 

A theory of weak interactions built around such a symmetry enjoys all the mathematical advantages of the Quantum Field Theories we have seen work so well for the electromagnetic and strong force, but it is immediately suspicious. How can a neutrino be interchanged with an electron, when they are so different when it comes to the electromagnetic charge? 

 And then, there is the problem of mass. We saw above that fermions cannot have a mass as only left-handed particles feel the weak force. We have now yet another problem, since the weak force is short range, and this means it is mediated by massive messengers, not the massless ones predicted by the symmetry. 
 
 \subsection{Symmetry Breaking and the Higgs boson}
 
 The crucial point is that although we do not {\it see} the symmetry, it does not mean that it does not exist. 
\begin{figure}[h]
\includegraphics[width=.5\columnwidth]{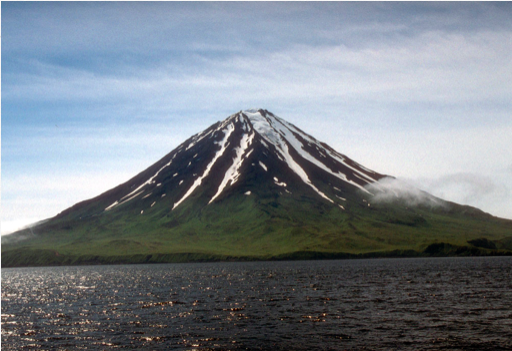}\includegraphics[width=.3\columnwidth]{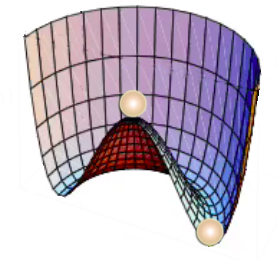}
\caption{Simmetries can be present but hidden to an observer  in a landscape, or to a particle in a potential. }
\label{fig-ssb}
\end{figure} 
Consider the landscape in the left panel of in Fig. \ref{fig-ssb}. It looks symmetrical to an observer standing on top of the volcano, the same all around, but highly asymmetrical to one sitting on the beach.  In the right panel, a particle is subject to a potential with a similar mountain-like shape. The symmetry is evident if the particle is at point zero, on the maximum, but this is not a stable place to be. The particle will fall to its state of lower energy, somewhere in the valley, and the symmetry will be hidden, or as is more often (if somewhat misleadingly) said, spontaneously broken. 

It was Yoichiro Nambu who  introduced the idea of spontaneous symmetry breaking to particle physics back in 1960~\cite{Nambu:1960xd}, with a stroke of genius that  revolutionized the field. Only a year later, Jeffrey Goldstone produced a fundamental work~\cite{Goldstone:1961eq} that elaborated this idea and transformed it into a full-fledged theory.  Nambu and Goldstone focused on the case of global, not local, continuous symmetries, but the essence is very similar. In their case, the result of symmetry breaking was the emergence of massless pseudo-scalar particles, the so-called Nambu-Goldstone bosons.  While Nambu was eventually rewarded the Nobel prize for his crowning achievement ("the discovery of the mechanism of spontaneous symmetry breaking in subatomic physics"), Goldstone's contribution was not acknowledged by the committee. Remarkably, Goldstone  noticed that in the process of symmetry breaking there must exist a scalar boson, with spin 0,  that becomes massive as it spontaneously acquires a non-zero value in the vacuum. 

Building on their work, Peter Higgs pointed out in 1964 that, when the symmetry is gauged, its spontaneous breakdown    makes the messenger gauge bosons massive \cite{Higgs:1964pj}. This is known as the Higgs mechanism. In the process, Higgs reproduced  Goldstone's  prediction of a massive scalar boson - known today as the Higgs boson. At the same time, Brout and Englert~\cite{Englert:1964et} independently discovered the Higgs mechanism, so the name is clearly unfair.   Englert and Higgs deservedly shared the Nobel prize   when the Higgs boson was finally discovered in 2012 (Brout had deceased meanwhile), but sadly Goldstone's original prediction was again overlooked by the Nobel committee.  So much about prizes.

%Spontaneously broken symmetries in field theory were first studied in the fundamental works by Nambu \cite{Nambu:1960xd} and Goldstone \cite{Goldstone:1961eq} for the case of global symmetries. Goldstone even noticed that in the process there must exist a scalar boson that becomes massive as it spontaneously acquires a non-zero value in the vacuum. Building on their work, Peter Higgs pointed out in 1964 that in the case of gauge symmetries, spontaneous symmetry breaking  would produce not only a massive scalar boson, now known as the Higgs particle, but that messenger bosons would also acquire a mass \cite{Higgs:1964pj}. 

Finally in 1967, in a seminal work, Weinberg incorporated~\cite{Weinberg:1967tq} the Higgs mechanism into  Glashow's $SU(2) \times U(1)$ gauge theory, creating thus a self-contained predictive theory of weak and electromagnetic phenomena. The Standard Model was born. 

\begin{figure}[h]
\includegraphics[width=.7\columnwidth]{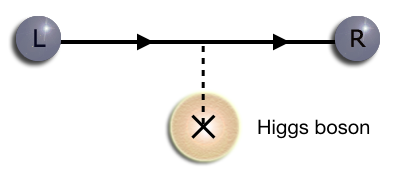}
\caption{The Higgs boson connects left and right particles, giving them a mass.}
\label{fig-higgs}
\end{figure}

Crucially, in Weinberg's theory the Higgs boson can also connect left and right particles as in Fig. \ref{fig-higgs}, and give them a mass. 

The vacuum is then filled with a Higgs field that solves all the problems with the mass without spoiling the symmetry. It  gives mass to the gauge bosons $W^+, W^-$ and $Z$, to the fermions, and to itself.  There is indeed a symmetry connecting particles with different charges (like electron and neutrino, or up and down quarks), but it is spontaneously broken, hidden.  The weak and electromagnetic interactions are treated together, into the unifying framework of electro-weak interactions.

In the elementary particle world, the Higgs boson (in a just world, it would be called Goldstone-Higgs-Weinberg boson) governs all question of mass. Coupling with almost every particle, filling the void with its presence,  the Higgs seems the perfect opposite of our aloof witness the neutrino. In his popular book, Leon Lederman calls it ``God's particle''.  He has said that he actually intended to call it the ``goddamn particle'',  only his editor would not have it. But in a sense both names do it justice: it plays a divine role when it comes to mass, but with no spin and no charge, weakly coupled to ordinary matter, it was terribly difficult to detect. It was discovered finally at the LHC in 2012, 45 years after it was invoked to make sense of the Standard Model and 48 years after its original conception. Talk about waiting.

Most important,  when the Higgs boson decays into a particle that gets its mass through the Higgs mechanism,  the Standard Model requires the amplitude of the process  to be proportional to that mass. This provides a key test to the theory, confirmed by increasing amounts of data from the LHC experiments. The origin of mass for charged fermions and gauge bosons seems clear: but what about the neutrino? In the Standard Model original picture (Fig. \ref{fig-SM}), there is no right-handed neutrino as a direct manifestation of its maximal left-right asymmetry. And it is precisely this asymmetry
that ensures the neutrino remains massless in spite of the Higgs mechanism. 

The issue of the existence of the right-handed neutrino is directly connected with the principle of minimality which lies at the core of the Standard Model. Minimality dictates that a particle with no interactions has no place in a model of particle interactions. If a right-handed neutrino is added, certainly the theory would look more symmetric -all other fermions have their right-handed counterparts. But all the predictions would be lost were one to give up minimality, as it always happens with theories of natural phenomena. With no rules on how to proceed, a Pandora box  would open up: what should be added and why? Add a right-handed neutrino just to get neutrino mass? There are numerous other possibilities, including much simpler ones. For this reason, the fathers of the Standard Model had stuck to its minimal version and proudly predicted neutrino to be massless.

We now know this is not true, neutrino is massive as all other fermions, if incomparably lighter, and this renders  the Standard Model incomplete. It is important to be clear that this is not a ``failure'':  the Standard Model is a greatly successful high precision theory in its domain, and as such cannot fail. But it needs a completion when it comes to neutrino mass, in the same sense as Newton's  theory of gravitation did not fail, but got completed by the theory of General Relativity.  
In building this completion,  the origin of neutrino mass becomes a burning question: is it the same Higgs mechanism that governs the mass of the electron? And equally important, what is the nature of neutrino mass: Dirac, Majorana, or both?

\section{A theory of neutrino mass}

\subsection{Left-Right symmetry}

\begin{figure}[h]
\includegraphics[width=.9\columnwidth]{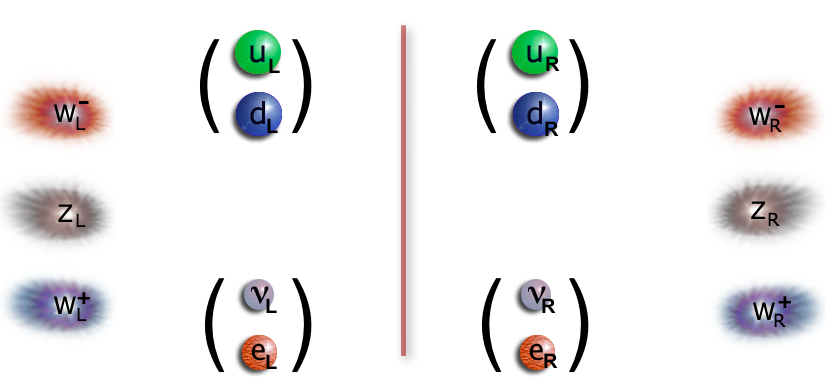}
\caption{The Left-Right symmetric model.}
\label{fig-leftright}
\end{figure}

Look again at the picture of the Standard Model in Fig. \ref{fig-SM}. Doesn't it bother you, this lack of left-right symmetry?  It bothered deeply one of the authors of this essay. Building on the work by Jogesh Pati, Abdus Salam and Rabi Mohapatra \cite{Pati:1974yy,Mohapatra:1974gc}, he and Mohapatra suggested in 1975 that parity was not just broken, but {\it spontaneously} broken \cite{Senjanovic:1975rk,Senjanovic:1978ev}. The left-right symmetry of the world is just hidden at larger energies, and the complete picture would look like Fig. \ref{fig-leftright}.

To the weak interaction felt only by left-handed particles, a new one is added felt only by right-handed particles. The new symmetry group, $SU(2)_R$, is the mirror image of the previous, now called $SU(2)_L$, and the new interaction has its own three messengers, $W^+_R, W^-_R$ and $Z_R$. Parity is broken spontaneously by the vacuum expectation value of other types of Higgs bosons, and those messengers become so heavy that we have not yet seen them. The new, right-handed weak force is felt by particles only at very high energies. 

And yes, there must be a right-handed neutrino $\nu_R$, forming a doublet with $e_R$. In the Left-Right theory, neutrinos were predicted to be massive long before the experiment said so. Moreover, both versions of mass, Majorana and Dirac, are possible, leading to an implementation of the so-called seesaw mechanism~\cite{Minkowski:1977sc, Mohapatra:1979ia, Yanagida:1979as, Glashow:1979nm, GellMann:1980vs},  as we now describe.

\subsection{ The see-saw mechanism}

\begin{figure}[h]
\includegraphics[width=.5\columnwidth]{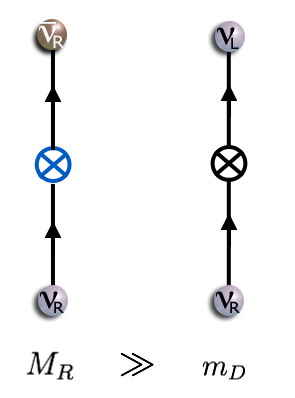}
\caption{Hierarchy of the Dirac and Majorana mass terms for a neutrino in the Left-Right theory}
\label{fig-numasses}
\end{figure}

In the Left-Right symmetric theory, a Majorana mass term for right-handed neutrinos comes as a result of spontaneous breaking of Parity at a high scale - corresponding to the mass of $W_R$, the right-handed messenger. A Dirac mass term, on the other hand, is provided, as in the Standard Model, by the Higgs vacuum expectation value, and it is proportional to the mass of $W_L$, much lower, as shown in Fig. \ref{fig-numasses}

These states with two kinds of masses are not what you expect of a particle, with a definite mass. What we call the neutrino is a linear combination of these states, and to find it we need the neutrino mass matrix  
\beq
\begin{array}{cc}
 & \begin{array}{cc}  \nu_R \; & \; \nu_L \end{array}  \\
\begin{array}{c} 
\nu_R  \\ \nu_L \end{array}
& \left( \begin{array}{cc} M_R & m_D \\ m_D & 0  \end{array} \right) \\
\end{array}
\eeq

The physical states are found by diagonalizing this matrix, leaving us with with two Majorana fermions. Working to the leading order in $m_D/M_R$, it is easy to see that the familiar light one $\nu$ has a small mass
\beqa\label{numass}
  m_\nu \simeq - m_D^2/M_R\,,
  \eeqa 
 precisely because there is a heavy one - we will call it $N$ - with a large mass 
 \beqa\label{Nmass}
 m_N \simeq M_R. 
 \eeqa
 Thus the name seesaw. 

The new neutrinos are approximately the two orthogonal combinations
\beqa\label{nuN}
\nu &=&  \,  \nu_L - \frac{m_D}{M_R} \,   \nu_R \nonumber \\
N &=& \frac{m_D}{M_R} \, \nu_L  +  \, \nu_R\,.
\eeqa
So the light neutrino is mostly the familiar $\nu_L$, while  the new $N$ is heavy and made up mostly of $\nu_R$ (clearly both the trace and the determinant of the neutrino mass matrix are preserved in the process).   

The picture is more complicated when all three generations of neutrinos are included, but the essence is the same. The simplicity of this picture, the insight it provides, and the richness of emerging physics has made seesaw the central scenario behind the smallness of neutrino mass. It implies what Majorana envisioned with his genius: both $\nu$ and $N$ and Majorana fermions, resulting in the violation of lepton number, both at low and high energies.

All it takes for it to work is a right-handed neutrino with  a large Majorana mass term and a small Dirac mass, however it is its implementation in  Left-Right theory that has both the predictive power and a physical origin: neutrino mass gets deeply connected with Parity violation. Simply, neutrino is light because parity is so strongly broken -- and as long as you don't assume that parity is never restored (as in the Standard Model),  neutrino becomes the door to new physics, as we shall see.

The seesaw mechanism was suggested independently by several groups, in theoretical frameworks ranging from  Left-Right symmetry~\cite{Minkowski:1977sc, Mohapatra:1979ia}, to grand unification of strong and electro-weak forces~\cite{Glashow:1979nm, GellMann:1980vs}, to the idea of family symmetry~\cite{Yanagida:1979as} of different generations. 

Today, however, the seesaw is often treated simply as an addition of right-handed neutrino $N$ to the SM.  While possible in principle, a great deal of the physical picture and (most important) the phenomenological predictions get lost this way. There is no real substitute for a full-fledged, predictive theory. 

There is more to be said about the seesaw, its variations and its relationship with the physics beyond the Standard Model, especially the Grand Unification of all known forces (without gravity). We will avoid these rather technical points and discuss only its salient features, and the connection with hadron colliders such as the LHC; the interested reader is referred to~\cite{Senjanovic:2011zz, Senjanovic:2011zf} for a review and further references.

\section{Neutrino as a probe of new physics}

\begin{figure}[h]
\includegraphics[width=.7\columnwidth]{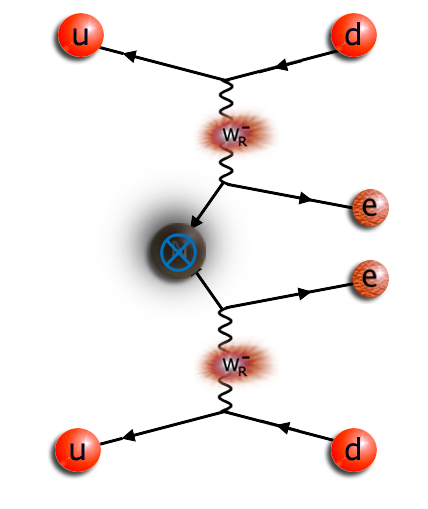}
\caption{The right-handed interaction  provides a different source for neutrinoless double beta decay in the Left-Right theory, through the heavy right-handed neutrino $N$.}
\label{fig-Ndoublebeta}
\end{figure}

If neutrinos were Majorana particles, there would be no conservation of lepton number, as we have said. Lepton-number-violating processes, such as neutrinoless double-beta decay,  can then shed light in the nature of neutrino.  The double exchange of the $W$ boson (or better, $W_L$ in Left-Right theory), with the neutrino Majorana mass does the job, and it is often said that neutrinoless double beta decay is thus the probe of neutrino Majorana nature. 

The situation, however, is more subtle as pointed out by Feinberg and Goldhaber more than 60 years ago~\cite{ Goldhaber:1959}, when they argued that some unknown new physics could be behind this process. With the advent of the Standard Model this was later made more precise: In their seesaw paper, Mohapatra and GS~\cite{Mohapatra:1979ia} noticed that a right-handed interaction with the new bosons $W_R$ and the heavy neutrino $N$ would naturally provide such a new source for neutrinoless double-beta decay, as in Fig. \ref{fig-Ndoublebeta}. 

In other words, neutrinoless double beta could play an even more important role, by allowing to probe the theory responsible for neutrino mass, instead of the mass itself. It is impossible to overemphasise this fundamental fact. And we could have a simple way of knowing what lies behind neutrinoless double beta decay: if at least one electron comes out right-handed, then it is new physics, period. If both were to end up left-handed, the situation would be more complex, and would require more work to untangle it. But this is a technical issue, beyond the scope of this review. 

\subsection{Lepton Number Violation at colliders: the Keung-Senjanovi\'c process}

\begin{figure}[h]
\includegraphics[width=.7\columnwidth]{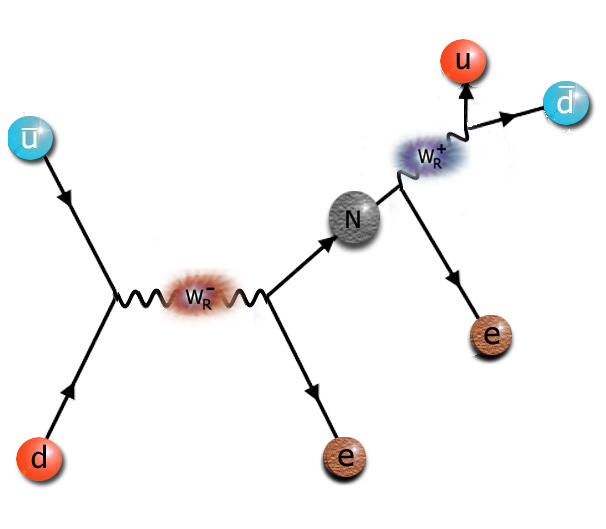}
\caption{The Keung-Senjanovi\'c process, representing lepton number violation at the hadron collider such as the LHC. It is a direct result of the Majorana nature of $N$. }
\label{fig-ks}
\end{figure}

The next step would end up being much more decisive. In 1983, it was  suggested by Keung and GS~\cite{Keung:1983uu} that the massive $N$  provides a very specific signature to look for in future colliders.
 It involves the scattering of proton and antiproton (proton), and produces two leptons of the same sign as a result, as shown in Fig. \ref{fig-ks}. The two quarks in the final states are what one calls jets, bunches of hadrons produced when the quarks - which are forever confined inside the hadrons - hadronise due to the strong interactions. 

It can be seen that it is a sort of rotation of the diagram  of neutrinoless double-beta decay in Fig. \ref{fig-Ndoublebeta} (recalling that changing the direction of the arrow amounts to change particle with antiparticle). The end result is the production of two same-sign leptons, electrons in the figure. We start with lepton number=0 and end up with 2, providing a very clear signature in the collider, in total analogy with the low energy neutrinoless double beta decay. There is a profound connection between the two processes~\cite{Tello:2010am},
 and both have the potential of being observed in near future. 

\begin{figure}[h]
\includegraphics[width=.7\columnwidth]{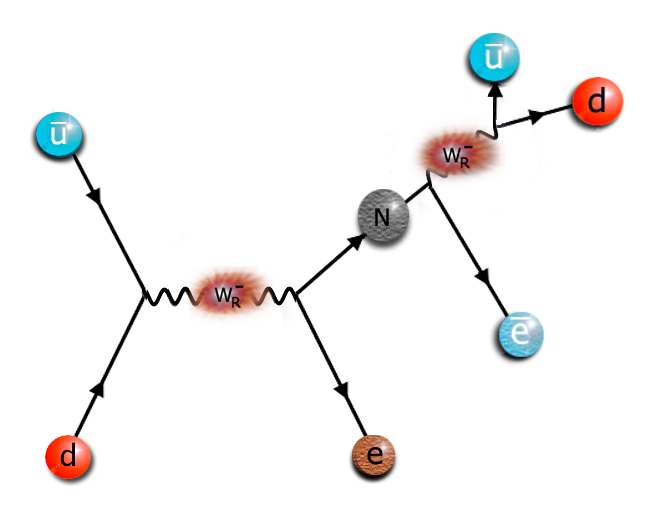}
\caption{A complementary version of the above process,  producing no lepton number violation. Since $N$ is Majorana fermion, the rate for this process is the same as for the lepton number violating one in Fig. \ref{fig-ks}. }
\label{fig-antiks}
\end{figure}

Since $N$ is a Majorana particle and therefore its own antiparticle, the process of Fig. \ref{fig-antiks} happens with the same strength. In that case, the end result is an  electron and  a positron. In this way, this process allows a unique probe of the Majorana nature of neutral leptons $N$ -- once produced they must decay equally into a charged lepton and into its anti-particle, finally offering a direct possibility of verifying the Majorana's theory. 

%Here we should stress a fundamental point. In the Left-Right symmetric theory $N$ is produced directly by the $W_R$ gauge boson, and the essence lies in the $N$ signatures, not how it is produced. In principle, even in hard practice, it could be also produced through a small mixing with the light neutrino (recall that they mix, as we saw in Eq.\ref{nuN}) and the $W$ production; whatever it is, the resulting physics is one and the same. One ends up with the  direct lepton number violation in the form of same sign charged leptons and at the same time equal rate for the opposite sign leptons, a crucial signature of the Majorana nature of $N$.

\subsection{A predictive, self-contained theory of neutrino mass}

Thus we see that the Left-Right symmetric model does much more than just accommodating a neutrino mass. It connects it with Parity violation in the weak interactions, and moreover creates a program for verifying the Higgs mechanism for the neutrino. This is similar to what the Standard Model does for charged fermions, according to the classic Weinberg 1967 work~\cite{Weinberg:1967tq} that is now being probed at the LHC. The Higgs bosons gives charged fermions their mass, and this implies that its decay rates into a fermion-antifermion pair are proportional to the charged fermions (and Higgs) masses
\beq
\Gamma(h \to f\bar f)  \propto m_h\, m_f^2\,,
\eeq
allowing for a precise identification of the new particle discovered in 2012 as the Higgs boson and a validation of the Standard Model. In this way, today we know for certain that the $W$ and $Z$ gauge bosons, and the third generation of quarks and leptons, the top and the bottom quark and the tau lepton, get the mass from the Higgs mechanism. 

In other words, the Standard Model, which started modestly as a gauge theory of weak interactions, over the years turned into a theory of the origin of mass. Mass is now a dynamical parameter whose value lies in one-to-one correspondence with an associated decay of the Higgs boson. This has been a remarkable achievement, and  we can say that the Standard Model is a self-contained, predictive theory of all fermion masses, save the neutrino one. This is why the origin of neutrino mass, as we have argued from the onset, is so important - it holds the key to a new fundamental theory beyond what we know today. And the Left-Right theory has all the ingredients to provide that.

The crucial point is that  Left-Right theory ends up doing for neutrino precisely what the Standard Model does for other fermions, namely it connects neutrino mass to a number of associated decays of both new scalars and the right-handed neutrino $N$, the latter being particularly important and illuminating. Through the mixing with the light neutrino, $N$ decays into the $W$ boson and the electron (or any charged lepton for that matter), and the decay rate is found to be proportional to its mass squared and the light neutrino mass~\cite{Nemevsek:2012iq,Senjanovic:2016vxw}
\beq
\Gamma(N \to W^+ \ell) \propto m_\nu\, m_N^2\,,
\eeq
where $\ell$ stands for any of the three generations charged leptons. Since $N$ is a Majorana fermion in this theory,   the rate of  decay into   anti-particles is predicted to be the same.
Thus the Majorana nature of $N$ can be directly tested, and in turn the  Majorana nature of $\nu$through the see-saw mechanism. 

This justifies the claim that we have been making from the beginning. 
It should be stressed that this is just one of many such examples~\cite{Senjanovic:2018xtu}. Left-Right theory does not just predict a neutrino mass, but connects it, in a calculable manner, with a plethora of new decays. Just like the Standard Model, it is a self-contained and predictive theory, a theory of the origin of neutrino mass. 

The way it works is as follows. If $N$ was to be discovered at the LHC, it would become possible to measure its mass and mixings with the charged leptons, thus making a clear prediction for neutrinoless double beta decay. And vice versa, suppose neutrinoless double beta decay is observed. If  the outgoing electron chirality could be measured, one would know whether the process is induced thorough $W$ and neutrino, or $W_R$ and $N$. A right-handed electron would strongly suggested the Left-Right theory as the culprit behind this process and encourage the hadron collider search for $N$. Since at the hadron collider it is possible to identify different charged leptons, clear predictions for  low energy lepton flavor violation processes (like $mu \to e \gamma$, $\mu \to e e \bar e$ and $\mu \to e$ conversion in nuclei) would follow suit.

\section{Epilogue}

Everything else aside, one crucial thing we learned about our aloof neutrino: its interactions with matter are extremely rare, but its interactions with the history of physics and the events that shaped our knowledge of elementary particles have been frequent and fruitful. For all its indifference, it has been a protagonist in the development of the Standard Model, with a key role on the meaning and origin of the concept of mass. We hope to have convinced the reader that neutrino mass, is a key to a door to new physics beyond the Standard Model. 

The neutrino  is the  only particle we know that can have both Dirac and Majorana versions of mass, so its study can shed a light on the nature of mass itself. Most important, the neutrino can be connected to the mystery of Parity violation in Nature, through the Left-Right symmetric model, a self-contained, complete theory of neutrino mass. As such, Left-Right theory has definite predictions that can be tested and linked to the nature of mass. 

The central role in this is played by the Keung-Senjanovi\'c process, a hadron collider production of the right-handed neutrino, with its subsequent decays into charged leptons and hadron jets. This leads to  direct Lepton Number Violation at colliders, and at the same time, allows to probe the Majorana nature of the right-handed neutrino through the predicted equality of its decay rates into charged leptons and their anti-particles. In turn, through the seesaw mechanism that gets untangled in the minimal Left-Right model, one can probe the origin of neutrino mass.

 It took a great deal of time to be able to  access the scale of Parity breaking required in Left-Right theory. In 1981, Beall, Bander and Soni \cite{Beall:1981ze} showed  that the mass of the right-handed $W_R$ boson needs to be larger than about 2 TeV in the minimal version of the theory, far above the accessible energies of the day.  Left-Right theory had to wait for the LHC, which at 13 TeV has made these energies available to experiment.
 
Today, both the CMS and ATLAS experiments at the LHC are  searching for the RH gauge boson, and have managed to set the limit $M_{W_R} \gtrsim 4-5 \,TeV$, depending on the mass of $N$~\cite{Aaboud:2018spl}.  The prospects for a possible LHC discovery are hopeful~\cite{Nemevsek:2018bbt}, and even more at the next hadron collider~\cite{Ruiz:2017nip}. 
Meanwhile, there has been important progress on the theoretical side: in particular the precise relation between  the quark mixings in the left-handed and right-handed sectors has been determined ~\cite{Senjanovic:2014pva,Senjanovic:2015yea}. This would allow clear experimental scrutiny at the LHC and equally clear predictions for low energy processes.

And so, almost half a century after its proposal,  Left-Right  symmetry is going through a renaissance from the theoretical point of view.  The origin of parity violation in weak interaction and with it, the origin of neutrino mass, has finally a chance of being probed experimentally. Perhaps Pauli is again proved right, and ``everything comes to him who knows how to wait''.

\section*{Appendix: Relativistic spinors}

In this Appendix we give a telegraphic overview of the relativistic notion of Dirac and Majorana spinors for the reader familiar with the $SU(2)$ group and the  fundamentals of Field Theory. 

\subsection*{Weyl spinors, Dirac and Majorana mass terms}

The first concept to introduce are the irreducible spin 1/2 representations of the Lorentz group. These are  two-component left- and right-handed chiral fermion Weyl fields $u_L$ and $u_R$, which transform under the Lorentz group as 
\beq  \label{lorentztrans}
u_{L,R} \to \Lambda_{L,R} \, u_{L,R}\,;\,\,\,\,\,\,\Lambda_L \equiv e^{i \vec\sigma/2 (\vec \theta + i \vec \phi)}\,,\,
\Lambda_R \equiv e^{i \vec\sigma/2 (\vec \theta - i \vec \phi)}\,,
\eeq
where $\sigma_i$ are the usual Pauli matrices.The three Euler angles $\vec \theta$ stand for rotations, and $\vec \phi$ denotes the boosts ($\tanh \phi_i = v_i $). The spinors $u_L$ and $u_R$ transform the same under the  rotations, but in an opposite manner under the boosts, hence the name left and right.
 
 The following bilinear combinations are easily shown to be Lorentz invariant 
 \begin{eqnarray}
 && u_L^T i \sigma_2 u_L \,;\,\,\,\,\,\,  \qquad u_R^T i \sigma_2 u_R \qquad({\rm Majorana \, type}) \nonumber \\
 &&u_L^\dagger u_R  \,;\,\,\,\,\,\,\,\,\,\,\,\,\,\,\,\,\,\,\,\,\,\,\,\,\, \qquad u_R^\dagger u_L \;\;\;\;\;\;\;\;\; \; \qquad({\rm Dirac \, type}) \,.
 \end{eqnarray}
% 
% Historically, the Dirac type came first, but in  a sense the Majorana invariant is even more fundamental for it needs only one species of spinors.
 These correspond respectively to Majorana and Dirac fermion mass terms, denoted as $m_M$ and $m_D$ hereafter.
 
 \subsection*{Dirac spinors}

 The second central concept is the Dirac algebra
 \beq
 \{ \gamma^\mu , \gamma^\nu \} = 2 g^{\mu\nu} \qquad g^{\mu\nu} = diag (1,-1,-1,-1)\,.
 \eeq
A convenient representation is
 \beq
\gamma^i = \left( \begin{array}{cc} 0 & \sigma^i \\ - \sigma^i &0  \end{array}\right) \, , \qquad 
\gamma^0 = \left( \begin{array}{cc} 0 & 1_2 \\  1_2&0  \end{array}\right) \,.
 \eeq

 With the introduction of the $\gamma_5$
 \beq
 \gamma_5 = i \gamma^1 \gamma^2 \gamma^3 \gamma^0  = \left( \begin{array}{cc} 1_2 & 0 \\  0 & - 1_2  \end{array}\right) \,;\,\,\{\gamma_5 ,\gamma_\mu\} = 0\,,
 \eeq
one can define L and R  projectors
 \beq
 P_{L.R} \equiv \frac{1 \pm \gamma_5}{2}.
 \eeq
 
A four component spinor $\psi$  transforms under the Lorentz group  as
\beq
\psi \to \Lambda \, \psi\,;\,\,\,\,\,\Lambda \equiv e^{i \Sigma_{\mu \nu}  \theta^{\mu \nu} }\,;\,\,\,\,\,\,\Sigma_{\mu \nu} = \frac{1}{4 i} [\gamma_\mu,\Gamma_\nu]\,,
\eeq
where $\Sigma_{\mu\nu}$ are the generators of the Lorentz algebra (in a reducible 4-dimensional representation).
%
%where 
%\beq
%\Sigma_{\mu \nu} = \frac{1}{4 i} [\gamma_\mu,\Gamma_\nu]
%\eeq
%
%generate Lorentz algebra and 
%
%where 
%\beq
%\Lambda \equiv e^{i \Sigma_{\mu \nu}  \theta^{\mu \nu} }\,;\,\,\,\,\,\,\Sigma_{\mu \nu} = \frac{1}{4 i} [\gamma_\mu,\Gamma_\nu]
%\eeq

 The $\Lambda_{L,R}$ introduced in  (\ref{lorentztrans}) are simply ($[ \gamma_5 , \Sigma_{\mu \nu}] = 0$ ),
  \beq
\Lambda_{L,R} = P_{L,R}  \Lambda\,.
\eeq
One can write
  \beq
\psi \equiv \psi_L + \psi_R\,;\,\,\,\,\,\psi_L = P_L \psi\,, \; \psi_R = P_R \psi\, ,
\eeq
or, in terms of the Weyl spinors
\beq
 \psi_L = \left( \begin{array}{c} u_L  \\  0 \end{array}\right) ,  \psi_R= \left( \begin{array}{c} 0  \\  u_R \end{array}\right)\,.
 \eeq
 The left-handed and right-handed projected spinors $\psi_{L,R}$ represent the chiral fermions, the cornerstone of our review.

 \subsection*{Dirac equation}
 
The propagation of the free Dirac spinor is described by the Dirac Lagrangian 
\begin{equation}
{\cal L} = i \overline \psi \gamma^\mu \partial_\mu \psi - m_D \overline \psi \psi \,.
\end{equation}
 from which one readily obtains the Dirac equation \eqref{dirac}.
 
 It is easy to see that the above Lagrangian posses a global phase transformation invariance 
\begin{equation}
\psi \to \exp {(i \alpha)} \psi\,,
\end{equation}
 which by Noether theorem leads to the conserved current 
\begin{equation}
j_\mu = \overline \psi \gamma_\mu \psi\,;\,\,\,\,\,\, \partial_\mu j^\mu = 0\,.
\end{equation}

 Gauge invariance simply means promoting $\alpha$ into a space-time dependent function $\alpha (x)$, where $x$ stands for the four vector $(t, \vec x)$. Clearly, in order for a theory to be invariant under  transformations that depend on the position of the observer, there must a messenger of this information.  This is seen manifestly from the gauge invariant from of the Dirac Lagrangian
\begin{equation}\label{covdirac}
{\cal L} = i \overline \psi \gamma^\mu D_\mu \psi - m_D \overline \psi \psi \,,
\end{equation}
 where 
\begin{equation}\label{covariant}
D_\mu = \partial_\mu - i e A_\mu\,.
\end{equation}
The gauge potential $A_\mu$  plays the role of a messenger, and this way gauge invariance is deeply associated with the beautiful concept of forces being dynamically transmitted.  $A_\mu$  transforms under a gauge transformation as
\begin{equation}
 A_\mu\ \to A_\mu + \frac{1}{e} \, \partial_\mu \alpha (x)
\end{equation}

 A learned reader will recognise the gauge invariance of Maxwell's electrodynamics, with the usual Maxwell Lagrangian 
\begin{equation}
{\cal L}_M =  - \frac{1}{4} F_{\mu, \nu} \,F^{\mu, \nu} \,,
\end{equation}
 providing the kinetic energy for $A_\mu$. 
 The anti-symmetric tensor $F_{\mu, \nu} = \partial_\mu A_\nu - \partial_\nu A_\mu$    is the electro-magnetic field
\begin{equation}
 F_{0 i} = E_i \,, \,\,\,\,\,\, F_{i j} = \epsilon_{i j k} B_k\,,
\end{equation}
 where $E_i$ and $B_i$ stand for the components of the electric and magnetic field, respectively.
 
 QED is the  Quantum Field Theory of Maxwell's classical electrodynamics, with $A_\mu$ representing the photon. 
 From \eqref{covdirac} and \eqref{covariant} one immediately obtains the photon interaction with the electron quoted in Section \ref{SMsection}:
 \beq 
{\cal H}_{QED} = e \bar \psi Q_{em} \gamma^\mu \psi A_\mu\,.
\eeq

 \subsection*{Parity and charge conjugation}
 
  The third important concept are the two discrete transformations, parity  and charge conjugation.
  
%  \begin{enumerate}
% 
% \item  
  \subsubsection*{Parity}
  
  Parity is the mirror symmetry interchanging left and right, defined as
\beq
 u_L \to u_R \qquad  u_R \to u_L\,.
\eeq
  So for Dirac spinors:
  \beq
  \psi \to \gamma_0 \psi.
  \eeq

   \subsubsection*{Charge conjugation}

% \item 
 The Dirac charge conjugation is defined through
 \beq
 C^T \gamma^\mu C = - \gamma_\mu^T \, , \,\,\,C^T = -C\,;\,\,\,\,\, (C =i \gamma_2 \gamma_0)\,,
 \eeq
(an explicit choice is $C = i  \gamma_2 \gamma_0$).
The charge-conjugated Dirac spinor is defined as
 \beq
 \psi^c \equiv C \overline{\psi}^T  \,.\hspace{2cm} 
 \eeq
 It is easy to show that 
\beq
\psi^c \to \Lambda \psi^c \; {\rm when} \; \psi \to \Lambda \psi\,.
\eeq

In other words, $\psi^c$ transforms the same way as $\psi$, i.e. it is also a proper spinor. The only difference between $\psi^c$ and $\psi$ lies in their charges being opposite when they get coupled to a gauge field such as the photon, so charge conjugation takes particles into anti-particles. 

Let us take a left-handed spinor 
\beq  \psi = \psi_L = \left( \begin{array}{c} u_L  \\ 0 \end{array}\right)\,. \eeq
It is easy to see that its charge conjugated spinor
\beq
\psi^c = \left( \begin{array}{c} 0 \\ -i \sigma_2 u_L^* \end{array}\right)\,,
 \eeq
is manifestly right-handed. In other words, charge conjugation is a mirror reflection, similar to parity, as we have been emphasising throughout.

%\end{enumerate}
 
The Majorana mass term defined above can be then written as
\beq
   \frac {1}{2} m_M (\psi_L^T C \psi_L + h.c. )\,,
\eeq
and the Dirac one as
 \beq
m_D (\overline \psi_L \psi_R +   \overline \psi_R \psi_L ) \equiv m_D \overline \psi \psi \,.
\eeq
%This in turn gives
%\beq
%\psi_D  = \left( \begin{array}{c} u_L\\u_R \end{array} \right)
%\eeq
%since
%\beq
%\psi_L  = \left( \begin{array}{c} u_L\\0 \end{array} \right) \;\;\;  and \;\;\; \psi_R  = \left( \begin{array}{c} 0\\u_R \end{array} \right)
%\eeq
%
%

It is convenient to work with left-handed antiparticles instead of right-handed particles
\beq
(\psi^C)_L \equiv C \bar \psi_R^T\,,
\eeq
in which case one can write a mass matrix for $\psi_L$ and $(\psi^C)_L$ in the Majorana notation $(\psi_1^T C \psi_2)$
\beq
\left(\begin{array}{cc}  
m_L & m_D \\
m_D & m_R\
\end{array}\right)\,,
\eeq
 where $m_L$ and $m_R$ are the Majorana mass terms of $\psi_L$ and $\psi_R$ respectively. The case of a pure Dirac fermion simply means $m_L = m_R = 0$, while $m_D = 0$ implies pure Majorana left-handed and right-handed spinors.

 \subsection*{Majorana spinors}

A four-component Majorana spinor $\psi_M$ is a sub-case of a Dirac spinor, with the particle identical to its own anti-particle. Mathematically, it is defined by $\psi_M = \psi_M^c$.
It is easy to show that $\psi_M$ can be written as
\beq\psi_M =  \left( \begin{array}{c} u_L  \\ -i\sigma_2 u_L^*  \end{array}\right)\,.\eeq
 In other words, the right-handed spinor $u_R$ is not an independent field as in the Dirac case - a Majorana fermion has half the degrees of freedom of a Dirac fermion.

The free Majorana Lagrangian takes the same form as the Dirac one
\begin{equation}
{\cal L} = i \overline \psi_M \gamma^\mu \partial_\mu \psi_M - m_M \overline \psi_M\psi_M\,,
\end{equation}
which facilitates computations for those familiar with  QED and Feynman techniques.
 This is equivalent to
\begin{equation}
{\cal L} = i u_L^\dagger  \sigma_-^\mu \partial_\mu u_L -  \frac{1}{2} m_M \left(u_L^T\,i\sigma_2 u_L + {\rm h.c.} \right)\,,
\end{equation}
where $\sigma^\mu_\pm = (I, \pm \sigma_i)$ is the analog of  the Dirac $\gamma$ matrices for the two-component Weyl spinors. 

It is now clear why the mass term $m_M$ is called Majorana mass term - it leads to a Majorana spinor. It is equally clear that the Majorana mass term breaks any possible charge associated with a Majorana fermion. Only an electrically neutral particle, such as the neutrino (and its right-handed analog $N$), can be a Majorana spinor. A further insight into the connection between charge conservation, QED, and Majorana fermions, is provided by the fact that the analog of the electro-magnetic current $j_\mu$ vanishes for a Majorana spinor $\overline \psi_M \gamma_\mu \psi_M = 0$. This reassures us that a Majorana 
fermion cannot carry electric charge, givings us confidence in the Majorana formalism.

\bibliography{nubbl}{}

\end{document}